\documentclass[12pt]{revtex4}
\usepackage[dvips]{graphicx}
\usepackage{delarray}
\usepackage{amsmath}
\usepackage{amssymb}
\usepackage{here}
\usepackage[usenames,dvipsnames]{xcolor}
\usepackage{url}
\usepackage[colorlinks=true,
		linkcolor=red,
		citecolor=blue,
		urlcolor=blue]{hyperref}
\usepackage{tabularx,ragged2e,booktabs}
\usepackage{bm}
\usepackage{rotating}
\usepackage{mathtools}
\usepackage{epstopdf}
\newcommand{\ord}{\mathcal{O}}
\newcommand{\be}{\begin{equation}} \newcommand{\ee}{\end{equation}}

\renewcommand{\thesection}{\Roman{section}}

\newcommand{\vpa}{v_{\|}}

\newcommand{\p}{\partial}

\newcommand{\new}[1]{\textcolor{black}{#1}}
\newcommand{\NewSecond}[1]{\textcolor{black}{#1}}
\newcommand{\NewThird}[1]{\textcolor{black}{#1}}

\newcommand{\sfincs}{{\sc SFINCS}}
\newcommand{\dkes}{{\sc DKES}}

\newcommand{\Ev}{\mathbf{E}}
\newcommand{\Bv}{\mathbf{B}}

\newcommand{\na}{\nabla}

\newcommand{\fullcircle}{\mbox{{\large$\bullet\,$}}} 
\newcommand{\fullsquare}{\mbox{\protect\rule[1.5pt]{5pt}{5pt}}}
\newcommand{\fulldiamond}{\protect\rotatebox{45}{\protect\rule[0.3pt]{5pt}{5pt}}}

\newcommand{\dashed}{\protect\mbox{\large -\!\! -\!\! -\!\! -}}

\newcommand{\broken}{\protect\mbox{-- -- --}}

\newcommand{\chain}{\protect\mbox{\large -\!\! $\cdotp$\!\! -}}
\newcommand{\longchain}{\protect\mbox{\large --\!\! $\cdotp$\!\! --}}

\newcommand{\full}{\protect\mbox{\large \textbf{------}}}

\newcommand{\fullcircleline}{\protect\mbox{\large \textbf{---$\!\!\!\bm \fullcircle \!\!\!$---}}}
\newcommand{\crossline}{\protect\mbox{\large \textbf{---$\!\!\!\bm \times \!\!\!$---}}}
\newcommand{\fullsquareline}{\protect\mbox{\large \textbf{---$\!\!\! \bm \fullsquare \!\!\!$---}}}
\newcommand{\fulldiamondline}{\protect\mbox{\large \textbf{--- \hspace*{-5mm} \fulldiamond \hspace*{-5mm} ---}}}

\newcommand{\plusline}{\protect\mbox{\large \textbf{---\protect\raisebox{0.5pt}{$\!\!\! \bm{+} \!\!\!$}---}}}
\newcommand{\starline}{\protect\mbox{\large \textbf{---\protect\raisebox{0.75pt}{$\!\!\! \bm{\star} \!\!\!$}---}}}

\definecolor{ColorZeff1p05Two}{HTML}{BBBB00}
\definecolor{ColorZeff1p05Four}{HTML}{444422}
\definecolor{ColorZeff1p05Five}{HTML}{BB0000}
\definecolor{ColorZeff1p05Six}{HTML}{FF9900}

\definecolor{ColorZeff2p0Three}{gray}{0.7}
\definecolor{ColorZeff2p0Four}{HTML}{CCCCAA}
\definecolor{ColorZeff2p0Five}{HTML}{22FF22}

\definecolor{NewCyan}{HTML}{00AAAA}
\definecolor{NewMagenta}{HTML}{AA00AA}

\begin{document}

\begin{center}
\Large
{\bf 
Impurities in a non-axisymmetric plasma: transport~and~effect on bootstrap current 
}\\
~\\*[0.5cm] \normalsize {A.~Moll\'en$^{1, 3, \dagger}$, M.~Landreman$^{2}$, H.~M.~Smith$^{3}$, S.~Braun$^{3, 4}$, P.~Helander$^{3}$
  \\
{\it\small $^1$ Department of Applied Physics, Chalmers University of Technology, G\"oteborg, Sweden}\\
{\it\small $^2$ Institute for Research in Electronics and Applied Physics, University of Maryland, College Park, Maryland 20742, USA}\\
{\it\small $^3$ Max-Planck-Institut f\"ur Plasmaphysik, 17491 Greifswald, Germany} \\ 
{\it\small $^4$ German Aerospace Center, Institute of Engineering Thermodynamics, Pfaffenwaldring 38-40, D-70569 Stuttgart, Germany} \\ \today \\
${}^{\dagger}$\href{mailto:albertm@chalmers.se}{albertm@chalmers.se}
}
\end{center}
\begin{abstract}
Impurities cause radiation losses and plasma dilution, and in stellarator plasmas the neoclassical ambipolar radial electric field is often unfavorable for avoiding strong impurity peaking. 
%
In this work we use a new continuum drift-kinetic solver, the \sfincs~code (the Stellarator Fokker-Planck Iterative Neoclassical Conservative Solver) [M.~Landreman et al., 
 \href{http://dx.doi.org/10.1063/1.4870077}{\em Phys. Plasmas} {\bf 21} (2014) 042503] which employs the full linearized Fokker-Planck\new{-Landau} operator, to calculate neoclassical impurity transport coefficients for a Wendelstein 7-X (W7-X) magnetic configuration. 
We compare 
\sfincs~calculations 
with 
theoretical asymptotes in the high collisionality limit. 
%
\new{We observe and explain a $1/\nu$-scaling of the inter-species radial transport coefficient at low collisionality, arising due to the field term in the inter-species collision operator, and which is not found with simplified collision models even when momentum correction is applied.} 
\NewSecond{However, this type of scaling disappears if a radial electric field is present.} 
We also use \sfincs~to 
analyze how the impurity content affects the neoclassical impurity dynamics and the bootstrap current. 
We show that a change in plasma effective charge $Z_{\mathrm{eff}}$ of order unity can affect the bootstrap current enough to cause a deviation in the divertor strike point locations.
\end{abstract}

\maketitle


\section{Introduction}\label{sec:Introduction}
3D plasma confinement concepts have an advantage over the tokamak, as they offer the potential of steady state plasma operation with no need for current drive \cite{fusionphysics12,helanderStellarator}. 
For steady state operation, impurity accumulation has to be avoided, since 
impurities cause plasma dilution, radiation losses and can lead to pulse termination by radiation collapse. 
The avoidance of impurity accumulation under relevant conditions is one of the most crucial tests for the capability of steady state 3D systems. 

In stellarators, particles can be trapped in helical magnetic wells and escape the plasma even in the absence of collisions. Thus the neoclassical transport is typically considerably larger than in axisymmetric configurations. 
In fact, at low collisionality (such as in a hot plasma core) neoclassical transport could be expected to dominate over the turbulent transport because of the $1 / \nu$-transport behavior, $\nu$ being the collision frequency \cite{helanderStellarator}. 
The flux-surface-averaged radial neoclassical particle flux of species $a$ can be written as
\begin{equation}
\left\langle \Gamma_a \cdot \na \psi \right\rangle \equiv \left\langle \int d^3 v \, f_{a1} \mathbf{v}_{d a} \cdot \na \psi \right\rangle = - n_a \sum_b \left[D_1^{ab} \left(\frac{d \ln n_b}{d \psi} + \frac{e_b}{T_b} \frac{d \Phi}{d \psi}\right) + D_2^{ab} \frac{d \ln T_b}{d \psi}\right],
\label{eq:NeoPartFlux}
\end{equation}
where $n_b$ is the density, $T_b$ is the temperature and $e_b \equiv Z_b e$ is the charge of species $b$ with $e$ being the proton charge, and the sum is taken over all plasma species. 
The brackets $\left\langle \ldots \right\rangle$ denote a flux-surface average. 
The $D_{j}^{a b}$:s are coefficients of the transport matrix, 
$\mathbf{v}_{d a}$ is the cross-field drift velocity and $f_{a1} = f_a - f_{Ma}$ is the departure from the Maxwellian part of the distribution function of species $a$. 
$\psi$ is a flux function representing a radial coordinate (often chosen to be the toroidal magnetic flux) and 
$\Phi$ is the electrostatic potential which relates to the radial electric field by $\displaystyle E_r = - d \Phi /d r$ where $r$ is the effective radius. 
The inter-species terms ($b \neq a$) are due to friction along the magnetic field 
\new{
between the different species 
}
\cite{helanderStellarator}. 
Moreover, in all stellarator collisionality regimes $D_2^{aa}$ is positive which implies that the temperature gradient drives an outward particle flux, tending to generate hollow density profiles. 

A consequence of collisionless trajectories not necessarily being confined is that different plasma species can have different radial transport rates. 
This results in a radial electric field $E_r$ to restore ambipolarity, which can be determined without knowledge of the turbulent transport since 
the radial neoclassical current is $1 / \rho_\ast = L / \rho_i$ ($\rho_i$ being the ion gyro radius and $L$ a typical macroscopic scale length) larger than the radial turbulent current unless $E_r$ is just right
 \cite{helanderNonaxisymmetric}. 
We note that the transport coefficients, $D_{j}^{a b}$, in Eq.~\eqref{eq:NeoPartFlux} depend on the value of $E_r$. 
The ambipolarity condition of the particle fluxes determining $E_r$ can have multiple roots, depending on the plasma parameters and the magnetic configuration. 
In the standard situation for stellarators, the neoclassical ambipolar radial electric field points radially inwards (i.e. a negative $E_r$), referred to as the ion root regime (in the electron root regime $E_r$ is instead positive). 
This electric field tends to cause impurity accumulation, which is particularly strong for heavy impurities whose charge numbers $Z$ are large. 
The electrostatic drive for impurity accumulation is not present in tokamaks or quasi-symmetric stellarators, where intrinsic ambipolarity implies that radial transport rates of each species must be independent of the radial electric field to high precision, 
and accumulation is merely caused by impurity-ion friction. 

In a single-impurity species plasma, we denote the species 
by the subscripts $e$ (electrons), $i$ (bulk hydrogen ions) and $z$ (impurities) respectively. 
For a trace impurity approximation (when $Z n_z / n_e \ll 1$) in a standard ion root plasma, the ambipolar electric field is essentially determined from the condition $\left\langle \Gamma_i \left(E_r\right) \cdot \na \psi \right\rangle = 0$ because $\left\langle \Gamma_i \left(E_r = 0\right) \cdot \na \psi \right\rangle  \gg \left\langle \Gamma_e \cdot \na \psi \right\rangle $ \cite{burhennNF2009}. 
Assuming $T_z = T_i = T$ and neglecting the inter-species coefficients in Eq.~\eqref{eq:NeoPartFlux}, 
it is possible to express the radial neoclassical impurity flux as 
\begin{equation}
\left\langle \Gamma_z \cdot \na \psi \right\rangle =  - n_z D_{1}^{zz} \left[\frac{d \ln n_z}{d \psi}  + \frac{D_{2}^{zz}}{D_{1}^{zz}} \frac{d \ln T}{d \psi} - Z \left( \frac{d \ln n_i}{d \psi } + \frac{D_{2}^{ii}}{D_{1}^{ii}} \frac{d \ln T}{d \psi}\right)\right] \equiv - D_{1}^{zz} \frac{d n_z}{d \psi} + n_z V_z,
\label{eq:ImpurityFluxEquation}
\end{equation}
where the term containing the factor $Z$ appears from substituting the ambipolarity condition for the bulk ions. 
$D_{1}^{zz}$ represents the diffusive part connected to gradients in $n_z$, while $V_z$ are the convective terms related to gradients of the bulk species. 
We note that Eq.~\eqref{eq:ImpurityFluxEquation} is also valid for an axisymmetric device, but the way to derive it differs from the way to do it for a stellarator because the particle flux in Eq.~\eqref{eq:NeoPartFlux} is independent of the radial electric field in axisymmetry. 
For tokamaks the coefficient $D_{2}^{ii}/D_{1}^{ii}$ in Eq.~\eqref{eq:ImpurityFluxEquation} can be negative indicating impurity screening. 
In stellarators however, earlier theory predicts it to be always positive and support impurity accumulation when the temperature profile is peaked \cite{burhennNF2009}. 
Typically a transient increase in impurity concentration is found, due to the inward convective impurity transport which drives impurity accumulation, until it balances the outward diffusive transport and $\left\langle \Gamma_z \cdot \na \psi \right\rangle = 0$. 
This results in a peaking of the impurity density profile with respect to the main ion profile, depending on the relative strength of the impurity pinch $V_z / D_{1}^{zz}$. 
This stationarity in the impurity profile is expected ultimately, even if the impurity core confinement time is large. 

Although neoclassical predictions often point towards impurity accumulation, 
there are experimental scenarios in stellarators where impurity accumulation has been avoided,  
e.g. in certain low-density scenarios at W7-AS and LHD with an outward radial electric field, 
or through the application of purification mechanisms such as radiofrequency heating. 
In LHD an extremely hollow profile of carbon impurity has been 
observed, 
referred to as an ``impurity hole'' \cite{idaImpurityHoleLHD,yoshinumaImpurityHoleLHD}. 
In W7-AS also a ``high density H-mode'' with low impurity confinement times has been discovered, 
accessible only through neutral beam injection \cite{McCormickPRL2002}. 
It is also possible that turbulent transport could significantly mitigate the neoclassical impurity accumulation. 
To enable the stellarator concept as a fusion reactor candidate with high pressure plasmas, the search for favorable scenarios capable of avoiding strong impurity accumulation is important.

In neoclassical theory, transport processes are usually assumed to be radially local and described by the linearized drift kinetic equation 
for the first order distribution function $f_{a 1}$ in $\rho_\ast$. 
To date, stellarator neoclassical calculations have predominantly been performed using simplified models for collisions. 
The most accurate linear operator available is the linearized Fokker-Planck\new{-Landau} operator \new{\cite{landau,RosenbluthPotentials} (also given in Appendix~\ref{sec:appendixA})}, which is used in a variety of axisymmetric calculations. 
However because of the extra dimension in stellarators, due to the lack of toroidal symmetry, stellarator calculations are more challenging 
and often only pitch-angle scattering collisions are retained (e.g. see Ref.~\cite{beidlerNF11}). This implies that coupling in the energy dimension is eliminated, and that 
momentum is generally not conserved. 
Momentum correction methods exist \cite{taguchiMomentum,sugamaMomentum,maassbergMomentum} for post-processing pitch-angle scattering results, but  these methods are not equivalent to using the full linearized Fokker-Planck\new{-Landau} operator. 
The difference between calculations with momentum-corrected pitch-angle scattering and full Fokker-Planck\new{-Landau} collisions could be expected to be particularly important for ion-impurity and impurity-ion collisions since the mass ratio is neither very large nor very small. 


In this work we study neoclassical impurity transport in stellarators using the continuum code \sfincs~(the Stellarator Fokker-Planck Iterative Neoclassical Conservative Solver), described in \cite{landremanSFINCS}. 
The code solves the radially local 4D drift-kinetic equation, retaining coupling in four of the independent phase space variables (two spatial and two in velocity). 
The code permits an arbitrary number of species, and it includes the linearized Fokker-Planck\new{-Landau} operator for self- and inter-species collisions, with no expansion made in mass ratio. 
\new{The present numerical implementation of this operator in the code is detailed in Appendix~\ref{sec:appendixA}.}
Our study will be restricted to a hydrogen plasma with one single impurity species. 
We emphasize that we will focus solely on neoclassical transport in this work, 
and 
stellarator turbulent transport is an area where much is still to be explored \cite{helanderStellarator}. 
Moreover, 
Ref.~\cite{regana} shows that the neoclassical impurity transport can be strongly affected by the variation of the electrostatic potential on a flux-surface, 
$\Phi_1 = \Phi - \left\langle \Phi\right\rangle$, which can be large enough to affect impurity species of high charge. 
\new{Although \sfincs~calculates $\Phi_1$, this effect is not included in the calculations presented here.} 

The remainder of the paper is organized as follows. 
In Sec.~\ref{sec:TransportCoefficients} we use \sfincs~to calculate neoclassical transport coefficients for the impurity particle flux, 
in a single-impurity-species hydrogen W7-X plasma. 
\NewSecond{We discuss the importance of the full linearized Fokker-Planck-Landau operator, by comparing to results from pitch-angle scattering calculations where momentum correction is applied afterwards.} 
Furthermore, we investigate in which collisionality regimes impurity screening from the pressure gradients can be expected. 
How the impurity content affects the neoclassical impurity dynamics and the bootstrap current in a non-axisymmetric device is then explored with \sfincs~in Sec.~\ref{sec:impurity}. 
In Sec.~\ref{sec:conclusions} we summarize the results and conclude.


\section{Impurity transport coefficients}\label{sec:TransportCoefficients} 
In this section we will calculate neoclassical transport coefficients $L_{j k}^{z b}$ for the impurity particle flux, 
which is written as a linear combination of the thermodynamic forces 
\begin{align}
	& \new{A_{z1}} = \frac{1}{n_z} \frac{d n_z}{d \psi} + \frac{Z e}{T_z} \frac{d \Phi}{d \psi} - \frac{3}{2 T_z} \frac{d T_z}{d \psi}, \nonumber \\
	& \new{A_{i1}} = \frac{1}{n_i} \frac{d n_i}{d \psi} + \frac{e}{T_i} \frac{d \Phi}{d \psi} - \frac{3}{2 T_i} \frac{d T_i}{d \psi}, \nonumber \\
	& A_{2}  = \frac{1}{T} \frac{d T}{d \psi} \; \new{\left(= A_{z2} = A_{i2}\right)}, 
\label{eq:ThermodynamicForces} 
\end{align}
where $2 \pi \psi$ from here on is specifically the toroidal magnetic flux. 
The motivation for our choice of the thermodynamic forces in Eq.~\eqref{eq:ThermodynamicForces} stems from Ref.~\cite{landremanSFINCS}, 
where the transport matrix for a single species drift-kinetic system of equations becomes Onsager symmetric if $E_r = 0$ and the forces are defined in this form. 
We assume a hydrogen plasma with a single impurity species present 
where the impurities and the main ions are in thermal equilibrium, $T_z = T_i = T$. 
We consider a $\mathrm{C}^{6+}$ impurity, as this species is expected to be the dominant impurity in W7-X. 
Note that we neglect the impurity-electron collisions, since their effect on the collisional impurity transport is negligible whenever 
$Z^2 n_z / n_i \gg \sqrt{m_e / m_i}$, which is practically always the case in reality \cite{helander02collisional}. 

\new{It is possible to express the impurity and ion fluxes in terms of the transport matrix $L_{j k}^{a b}$ defined as follows:
\begin{equation}
\frac{\iota \left(G + \iota I\right)}{c \, G}
\left(\begin{array}{c}
	 \frac{Z e v_z^{1/2}}{ T_z n_z^{1/2}}\left\langle \int d^3 v \, f_z \mathbf{v}_{d z} \cdot \na \psi \right\rangle \\ 
	\frac{Z e v_z^{1/2}}{ T_z n_z^{1/2}}\left\langle \int d^3 v \, f_z \frac{m_z v^2}{2 T_z} \mathbf{v}_{d z} \cdot \na \psi \right\rangle \\
	\frac{e v_i^{1/2}}{ T_i n_i^{1/2}}\left\langle \int d^3 v \, f_i \mathbf{v}_{d i} \cdot \na \psi \right\rangle \\
	\frac{e v_i^{1/2}}{ T_i n_i^{1/2}}\left\langle \int d^3 v \, f_i \frac{m_i v^2}{2 T_i} \mathbf{v}_{d i} \cdot \na \psi \right\rangle
\end{array}
\right) = \frac{G c}{\iota B_0}
\left(\begin{array}{cccc}
	L_{11}^{zz} & L_{12}^{zz} & L_{11}^{zi} & L_{12}^{zi} \\ 
	L_{21}^{zz} & L_{22}^{zz} & L_{21}^{zi} & L_{22}^{zi} \\
	L_{11}^{iz} & L_{12}^{iz} & L_{11}^{ii} & L_{12}^{ii} \\
	L_{21}^{iz} & L_{22}^{iz} & L_{21}^{ii} & L_{22}^{ii}
\end{array}
\right)
\left(\begin{array}{c}
	\frac{T_z n_z^{1/2}}{Z e v_z^{1/2}} A_{z1} \\ 
	\frac{T_z n_z^{1/2}}{Z e v_z^{1/2}} A_{z2} \\
	\frac{T_i n_i^{1/2}}{e v_i^{1/2}} A_{i1} \\
	\frac{T_i n_i^{1/2}}{e v_i^{1/2}} A_{i2}
\end{array}
\right).
\label{eq:TransportMatrix}
\end{equation}
The normalization is similar to Eq.~(40) of \cite{landremanSFINCS}, and implies that the matrix elements are dimensionless. 
Moreover, 
the matrix elements in Eq.~\eqref{eq:TransportMatrix} depend on $E_r$,
and it is possible to show that $L_{j k}^{a b} \left(E_r\right) = L_{k j}^{b a} \left(- E_r\right)$ when $T_i=T_z$. 
For a stellarator-symmetric magnetic geometry the elements are independent of the sign of $E_r$, $L_{j k}^{a b} \left(E_r\right) = L_{j k}^{a b} \left(- E_r\right)$. 
Hence, written in this form 
the matrix exhibits Onsager symmetry, i.e. $L_{j k}^{a b} = L_{k j}^{b a}$. 
\NewSecond{We note that the Onsager symmetry is true for both the ``full particle trajectories'' described by Eq.~(17) and the ``\dkes~particle trajectories'' described by Eq.~(18) in \cite{landremanSFINCS}.  
If we were to include additional matrix elements in Eq.~\eqref{eq:TransportMatrix} corresponding to bootstrap current and Ware pinch, the Onsager symmetry would generally not be fulfilled for the ``full particle trajectories'' when $E_r \neq 0$.} }
\NewThird{In this work the ``full particle trajectories'' is the default in the \sfincs~calculations. However, we will also present results from \sfincs~calculations with ``\dkes~particle trajectories'' for $E_r \neq 0$ (when $E_r = 0$ the two different models for the particle trajectories are equal).
}

\new{In this work we focus on the impurity particle transport,} 
and write
\begin{equation}
\frac{Z e \new{\iota} \left(G + \iota I\right)}{n_z c \new{T} G}\left\langle \Gamma_z \cdot \na \psi \right\rangle = 
\frac{G \new{T} c}{Z e \new{\iota} B_0 v_z}
\left\{
\new{\tilde{L}_{11}^{zz}} 
\new{A_{z1}} 
+ \new{\tilde{L}_{11}^{zi}} 
\new{A_{i1}} 
+ 
\new{\tilde{L}_{12}^{z}}
A_{2}
\right\}.
\label{eq:ImpFluxSFINCS}
\end{equation}
\new{Comparing Eqs.~\eqref{eq:TransportMatrix} and \eqref{eq:ImpFluxSFINCS} 
we can identify $\displaystyle \tilde{L}_{11}^{zz} = L_{11}^{zz}$, $\displaystyle \tilde{L}_{11}^{zi} = L_{11}^{zi} Z \sqrt{v_z n_i / v_i n_z}$ and $\displaystyle \tilde{L}_{12}^{z} = L_{12}^{zz} + L_{12}^{zi} Z \sqrt{v_z n_i / v_i n_z}$.}
\new{In Eqs.~\eqref{eq:TransportMatrix} and \eqref{eq:ImpFluxSFINCS}, $c$ is the speed of light in vacuum and $v_a = \sqrt{2 T_a / m_a}$ is the thermal speed of species $a$} 
(note that we employ Gaussian units). 
\new{To facilitate comparisons between different models, we will perform a study at $E_r = 0$ in this section but we will also use a more realistic finite value. 
}
The quantities $B_0$, $G$, $I$ and $\iota$ in \new{Eqs.~\eqref{eq:TransportMatrix}-\eqref{eq:ImpFluxSFINCS}} 
stem from the magnetic geometry specified in Boozer coordinates $\theta$ and $\zeta$ in which 
\begin{equation}
\mathbf{B} = K \left(\psi, \theta, \zeta\right) \na \psi + I \left(\psi\right) \na \theta + G \left(\psi\right) \na \zeta.
\label{eq:MagneticFieldBoozer}
\end{equation}
$B_0$ is the $\left(0, 0\right)$ Fourier mode amplitude of $B \left(\theta, \zeta\right)$, $c I / 2$ is the toroidal current inside the studied flux surface, 
$c G / 2$ is the poloidal current outside the flux surface and $\iota$ is the rotational transform \cite{helanderNonaxisymmetric}. 
In a typical stellarator, $\left|I\right| \ll \left|G\right|$ and $G \approx B_0 R$, where $R$ is the major radius of the device. 
Furthermore, in Boozer coordinates 
\begin{equation}
\left\langle X \right\rangle = \left(\int_0^{2\pi} d \theta \int_0^{2\pi} d \zeta \frac{X}{B^2}\right) / \left(\int_0^{2\pi} d \theta \int_0^{2\pi} d \zeta \frac{1}{B^2}\right).
\label{eq:FluxSurfAvgBoozer}
\end{equation}
We 
study a W7-X vacuum configuration with \new{$\iota = - 1$} 
at the plasma edge. 
For this geometry,
the normalization factors of Eq.~\eqref{eq:ImpFluxSFINCS} are defined such that they are 
\new{all positive except $\iota$ and $I$ (which have the same sign)}, 
and a positive flux is directed outwards whereas the density and temperature gradients in Eq.~\eqref{eq:ThermodynamicForces} are negative in the usual situation. 
The impurity transport coefficients, $\new{\tilde{L}_{11}^{zz}}$, $\new{\tilde{L}_{11}^{zi}}$ and $\new{\tilde{L}_{12}^{z}}$, are obtained with \sfincs, solving 
coupled linear drift-kinetic equations for each species 
with three different right-hand-sides. 
Similarly to Ref.~\cite{landremanSFINCS} we calculate the transport coefficients in terms of a normalized collisionality for the impurities
\begin{equation}
\nu_{z}' \equiv \frac{\left(G + \iota I\right) \nu_{zz}}{v_z B_0},
\label{eq:nuPrimeZ}
\end{equation} 
where 
\begin{equation}
\nu_{zz} =  \frac{4 \sqrt{2 \pi} n_z Z^4 e^4 \ln \Lambda}{3 m_z^{1/2} T_z^{3/2}}.
\label{eq:nuzz}
\end{equation} 
\new{The collision operator implemented in \sfincs~is discussed in Appendix~\ref{sec:appendixA}.}
In W7-X, the normalized collisionality can be expected to range from $\nu_{z}' \sim 0.01$ in the core of a high-temperature, low-density, low-$Z_{\mathrm{eff}}$ plasma to $\nu_{z}' \sim 10$ at the edge of a low-temperature, high-density, high-$Z_{\mathrm{eff}}$ plasma. 
In our calculations we will include unrealistically high values of $\nu_{z}'$ to be able to compare our results with analytic theory which is only available at high collisionality 
\NewSecond{(see Appendix~\ref{sec:appendixB})}. 
It is interesting to note that collisions among the impurity ions themselves are more important than collisions with the bulk ions even if $Z_{\mathrm{eff}}$ is not far above unity. This is because the ratio of the pitch-angle-scattering frequency between impurities and bulk ions $\nu_D^{zi}$ and the pitch-angle-scattering frequency between impurity ions $\nu_D^{zz}$ scales as
$\nu_D^{zi} / \nu_D^{zz} \sim n_i \sqrt{m_i}/ \left(Z^2 n_z \sqrt{m_z}\right)$ \cite{helander02collisional}. 
Thus, as soon as $Z_{\mathrm{eff}} - 1$ significantly exceeds $\sqrt{m_i/m_z}$, $z$-$z$ collisions are more important than $z$-$i$ collisions. 
For a hydrogen plasma with $\mathrm{C}^{6+}$ impurities, $\sqrt{m_i/m_z} = 0.29$. 

\new{We also define a normalized electric field similar to Ref.~\cite{landremanSFINCS},
\begin{equation}
E_{\ast} = \frac{c G}{\iota v_z B_0} \frac{d \Phi}{d \psi}.
\label{eq:NormalizedElectricField}
\end{equation}
This is the electric field normalized by the so-called resonant electric field. In axisymmetry it corresponds to the poloidal Mach number.} 

\NewSecond{
\subsection*{Simulation results at $E_r = 0$}\label{subsec:SimulationResults}
}

\begin{figure}[!ht]
\begin{center}
\includegraphics[width=1.0\textwidth]{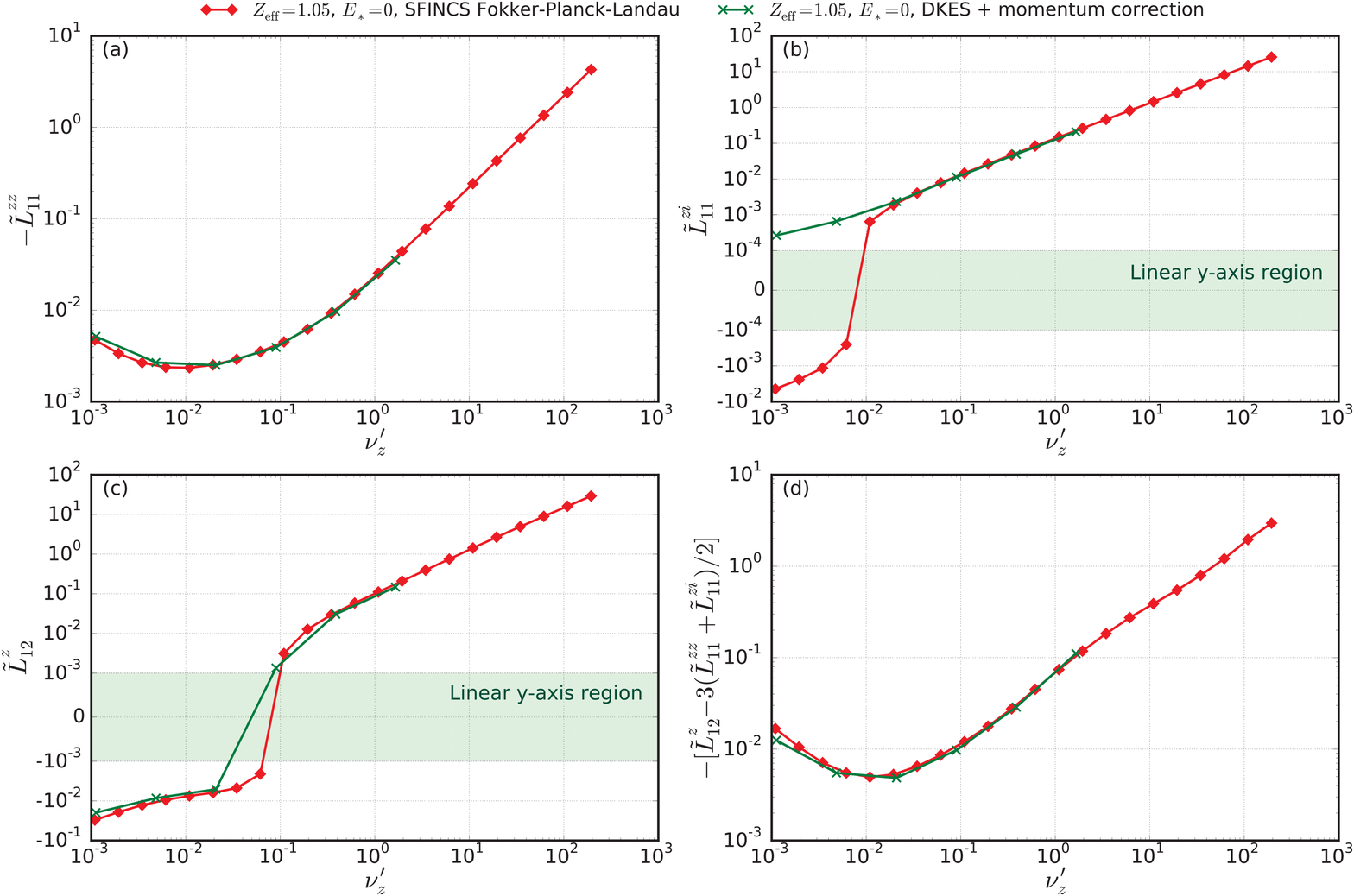}
\end{center}
\caption{Carbon ($Z = 6$) transport coefficients $\new{\tilde{L}_{11}^{zz}}$ (a), $\new{\tilde{L}_{11}^{zi}}$ (b), $\new{\tilde{L}_{12}^{z}}$ (c) and the temperature gradient coefficient $\new{\tilde{L}_{12}^{z}} - 3 \left(\new{\tilde{L}_{11}^{zz}} + \new{\tilde{L}_{11}^{zi}}\right)/2$ (d) as functions of normalized collision frequency $\nu_{z}'$ for a W7-X geometry at $\new{E_{\ast}} = 0$ and with $Z_{\mathrm{eff}} = 1.05$. 
\NewSecond{\sfincs~computations with the Fokker-Planck-Landau collision operator (\textcolor{Red}{\fulldiamondline}) 
are compared to \dkes~computations with momentum correction applied afterwards 
(\textcolor{Green}{\crossline}).
}
Note the double-logarithmic scale in (b)-\NewSecond{(c)}. 
}
\label{fig:CarbonTransportCoefficientsW7XZeff1p05}
\end{figure}

\begin{figure}[!ht]
\begin{center}
\includegraphics[width=1.0\textwidth]{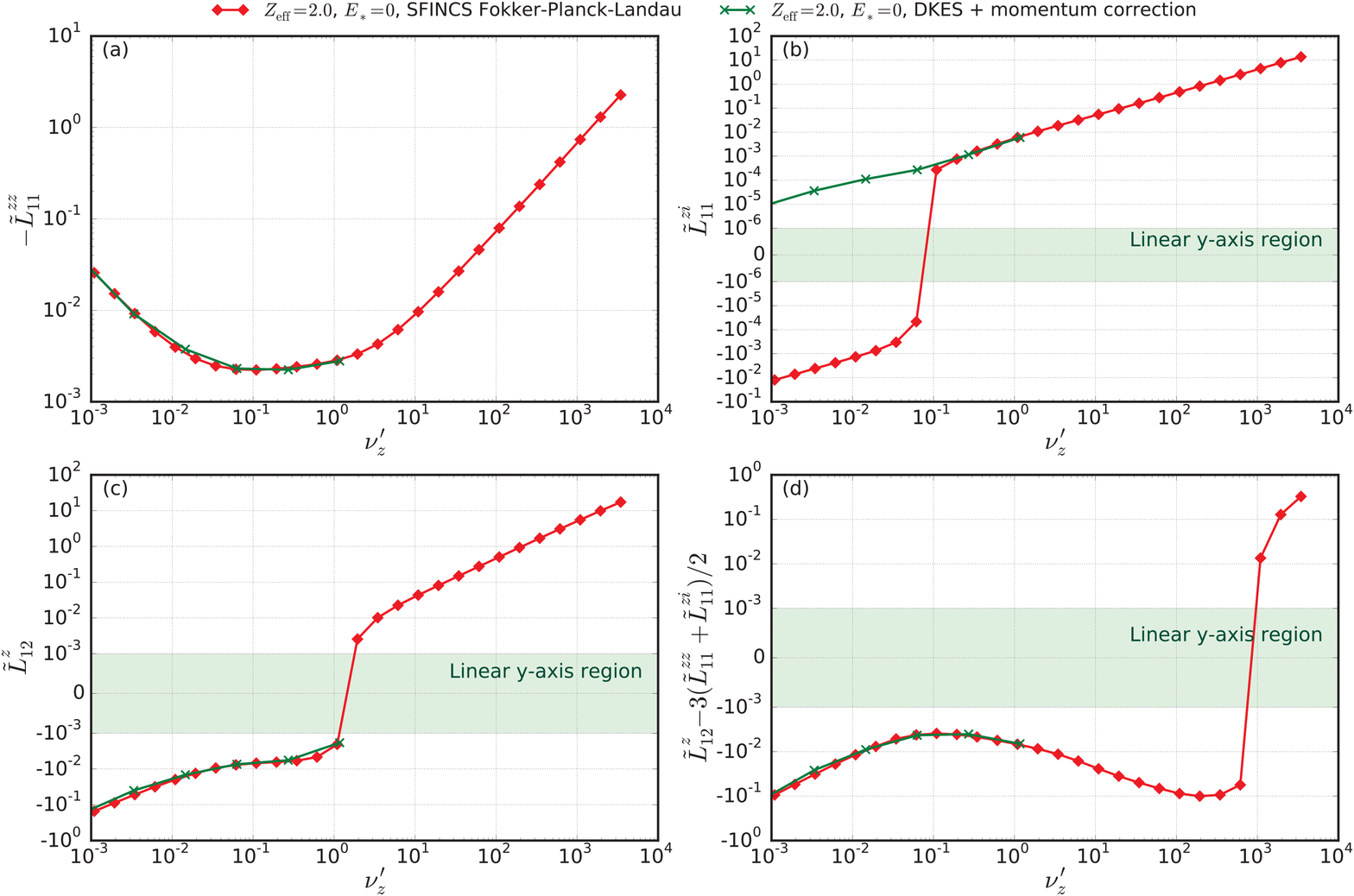}
\end{center}
\caption{Carbon ($Z = 6$) transport coefficients $\new{\tilde{L}_{11}^{zz}}$ (a), $\new{\tilde{L}_{11}^{zi}}$ (b), $\new{\tilde{L}_{12}^{z}}$ (c) and the temperature gradient coefficient $\new{\tilde{L}_{12}^{z}} - 3 \left(\new{\tilde{L}_{11}^{zz}} + \new{\tilde{L}_{11}^{zi}}\right)/2$ (d) as functions of normalized collision frequency $\nu_{z}'$ for a W7-X geometry at $\new{E_{\ast}} = 0$ and with $Z_{\mathrm{eff}} = 2.0$. 
\NewSecond{\sfincs~computations with the Fokker-Planck-Landau collision operator (\textcolor{Red}{\fulldiamondline}) 
are compared to \dkes~computations with momentum correction applied afterwards 
(\textcolor{Green}{\crossline}).
}
Note the double-logarithmic scale in (b)-(d). 
}
\label{fig:CarbonTransportCoefficientsW7XZeff2p0}
\end{figure}

\NewSecond{
In Figs.~\ref{fig:CarbonTransportCoefficientsW7XZeff1p05} and \ref{fig:CarbonTransportCoefficientsW7XZeff2p0} 
}
the carbon transport coefficients as functions of $\nu_{z}'$ for the W7-X standard configuration geometry at $r/a = 0.88$ are shown. 
$r$ is the effective radius related to the flux label through $\psi_N \equiv \psi / \psi_a = \left(r/a\right)^2$, where $a = 0.51~\mathrm{m}$ is the outermost effective minor radius and $\psi_a = \psi\left(\psi_N = 1\right)$.
At this radius the magnetic geometry parameters are $B_0 = 3.1~\mathrm{T}$, $G = 17.9~\mathrm{T m}$, $I = \new{-} 6.5 \times 10^{-7}~\mathrm{T m}$ and $\iota = \new{-} 0.93$. 
The minimum resolution used in the \sfincs~runs is $N_\theta = 17$, $N_\zeta = 49$ grid points in the poloidal and toroidal direction (per identical segment of the stellarator, where W7-X has a five-fold symmetry in the toroidal direction), $N_x = 5$ grid points in energy ($x = v / v_a$ with $v_a$ being the thermal speed of the species $a$) and $N_\xi = 24$ Legendre polynomials to represent the distribution function (here $\xi=v_\|/v$), and $N_L = 4$ Legendre polynomials to represent the Rosenbluth potentials.
Note however that the required resolution depends on the collisionality regime. At low collisionality $N_\zeta$ and $N_\xi$ both typically need to be larger than 100, because of the presence of an internal boundary layer between the trapped-passing boundary. 
At high collisionality instead $N_x$ 
typically 
has 
to be increased.

Results are presented for \sfincs~simulations with full linearized Fokker-Planck\new{-Landau} collisions, 
and for $Z_{\mathrm{eff}} = 1.05$ \new{at $E_{\ast} = 0$} (see Fig.~\ref{fig:CarbonTransportCoefficientsW7XZeff1p05}) and $Z_{\mathrm{eff}} = 2.0$ \new{at $E_{\ast} = 0$} (see Fig.~\ref{fig:CarbonTransportCoefficientsW7XZeff2p0}). 
\NewSecond{Note that plots (b)-(d) sometimes use a double-logarithmic vertical scale, since these transport coefficients can have either sign.}
The results are compared to simulations with the \dkes~(Drift Kinetic Equation Solver) code \cite{hirshmanDKES,vanRijDKES}. 
In contrast to \sfincs, 
\dkes~has 3 rather than 4 coupled phase-space coordinates, because energy coupling is neglected when solving the drift-kinetic equation. 
\dkes~employs pitch-angle scattering collisions, but momentum correction can be applied afterwards \cite{maassbergMomentum}. Here we use the momentum correction approach described in Ref.~\cite{taguchiMomentum}. 
\new{This approach is appropriate for relatively collisionless cases, because the energy scattering part of the collision operator only contains the first order Legendre components of the distribution functions. 
Moreover, in the parallel particle and heat flows the approach neglects terms corresponding to the Pfirsch-Schl\"uter form of the poloidal $\Ev \times \Bv$-drift. Consequently we will only show results at low collisionality 
($\nu_{z}' \lesssim 1$)  
using this momentum correction technique.} 
\NewSecond{We note that \dkes~employs different effective particle trajectories than \sfincs~(compare Eqs.~(17) and (18) in \cite{landremanSFINCS}), but the difference vanishes when $E_{\ast} = 0$.} 
In the short-mean-free-path limit, $\nu_{z}' \gg 1$, the impurity transport coefficients can be computed analytically in terms of the parallel current. 
The details are given in Appendix~\ref{sec:appendixB} and based on the theory presented in Ref.~\cite{braun}.
We note that \sfincs~retains $\Ev \times \Bv$-precession when solving the drift-kinetic equation for $f_{a1}$ \cite{landremanSFINCS}, although according to the formal ordering it should appear first at 
next order. 
\NewSecond{Since $\Ev \times \Bv$-precession is not included in the analytical high-collisionality calculation in Ref.~\cite{braun}, we only compare \sfincs~results to this high-collisionality theory at $E_{\ast} = 0$.} 
\NewSecond{The high-collisionality asymptotes for the W7-X case are shown in Figs.~\ref{fig:CarbonTransportCoefficientsW7XZeff1p05PfirschSchluter} and \ref{fig:CarbonTransportCoefficientsW7XZeff2p0PfirschSchluter} in Appendix~\ref{sec:appendixB}.}
These theoretical limits conform well with the \sfincs~computations in the appropriate limit. 

For both $\new{\tilde{L}_{11}^{zz}}$ and $\new{\tilde{L}_{12}^{z}}$ 
the momentum correction technique captures well the intrinsic momentum conservation of the full linearized Fokker-Planck\new{-Landau} operator. 
%
Interestingly, the \dkes~+ momentum correction curves fail to predict the sign change and $1 / \nu_{z}'$-scaling of $\new{\tilde{L}_{11}^{zi}}$ observed at low collisionality in the \sfincs~curves with Fokker-Planck\new{-Landau} collisions of both Figs.~\ref{fig:CarbonTransportCoefficientsW7XZeff1p05}~(b) and \ref{fig:CarbonTransportCoefficientsW7XZeff2p0}~(b), \new{i.e. at $E_{\ast} = 0$}. 

\begin{figure}[!ht]
\begin{center}
\includegraphics[width=0.7\textwidth]{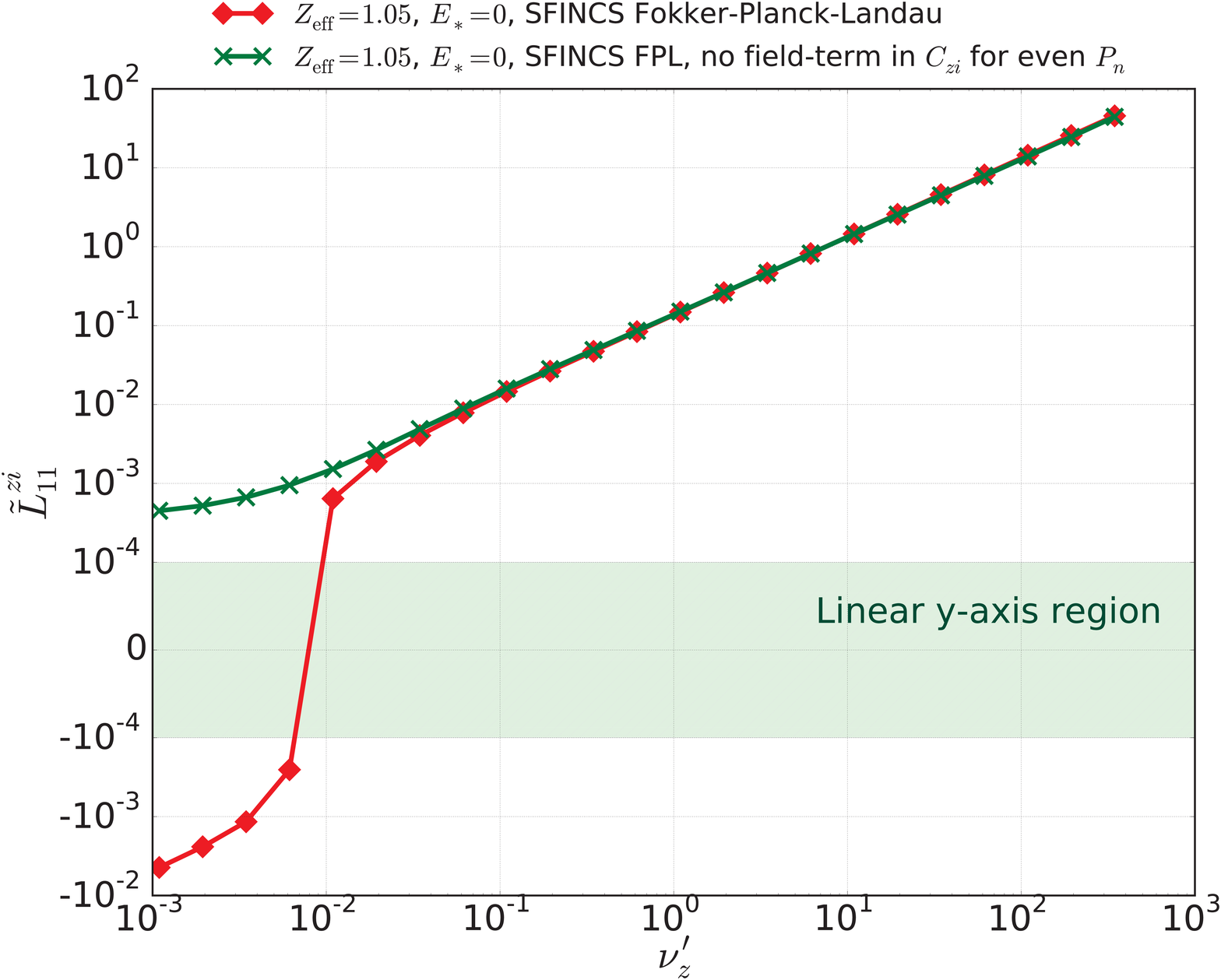}
\end{center}
\caption{Comparison of the $\new{\tilde{L}_{11}^{zi}}$ coefficient as function of normalized collision frequency $\nu_{z}'$ calculated with \sfincs~using full linearized Fokker-Planck\new{-Landau} collisions and when the even Legendre polynomials in the field term of the impurity-ion collision operator $C_{zi}$ have been suppressed, for a W7-X geometry at $\new{E_{\ast}} = 0$:  
Full linearized Fokker-Planck\new{-Landau} operator and $Z_{\mathrm{eff}} = 1.05$ 
(\textcolor{Red}{\fulldiamondline}), 
even Legendre polynomials in field term of $C_{zi}$ suppressed and $Z_{\mathrm{eff}} = 1.05$ 
(\textcolor{Green}{\crossline}). 
}
\label{fig:L11ziComparison}
\end{figure}

With Fokker-Planck\new{-Landau} collisions \new{at $E_{\ast} = 0$}, 
all impurity transport coefficients show trends of $1/\nu_{z}'$-transport at low collisionality and are proportional to $\nu_{z}'$ at high collisionality. 
Earlier work 
had assumed 
that the inter-species transport coefficients should be negligible compared to the self-species coefficients at low collisionality since the species 
interact via collisions 
\cite{helanderStellarator}. This is not in agreement with what we find in our \sfincs~calculations \new{at $E_{\ast} = 0$} as shown in Figs.~\ref{fig:CarbonTransportCoefficientsW7XZeff1p05} and \ref{fig:CarbonTransportCoefficientsW7XZeff2p0}, where the Fokker-Planck\new{-Landau} curves of $\new{\tilde{L}_{11}^{zi}}$ also exhibit $1/\nu_{z}'$-transport at low collisionality. 
This phenomenon is explained 
as follows. 
Since $\new{\tilde{L}_{11}^{zi}}$ is found by setting all thermodynamic forces to zero except the main ion gradient, the relevant impurity kinetic equation is 
\begin{equation}
\vpa \nabla_\| f_{z1}  = C_{zz} \left[f_{z1}, f_{Mz}\right] + C_{zz} \left[f_{Mz}, f_{z1}\right] + C_{zi} \left[f_{z1}, f_{Mi}\right] + C_{zi} \left[f_{Mz}, f_{i1}\right],
\label{eq:DriftKineticEqL11zi}
\end{equation}
where $C_{ab} \left[f_{a}, f_{b}\right]$ is the Fokker-Planck\new{-Landau} collision operator for species $a$ with distribution $f_{a}$ colliding with species $b$ of distribution $f_{b}$. 
At low collisionality the particle transport is carried by the trapped particles. 
If we perform bounce-averaging we annihilate the streaming term in Eq.~\eqref{eq:DriftKineticEqL11zi} and obtain 
\begin{equation}
0  = \overline{C_{zz} \left[f_{z1}, f_{Mz}\right]} + \overline{C_{zz} \left[f_{Mz}, f_{z1}\right]} + \overline{C_{zi} \left[f_{z1}, f_{Mi}\right]} + \overline{C_{zi} \left[f_{Mz}, f_{i1}\right]},
\label{eq:DriftKineticEqL11ziBounce}
\end{equation}
where the bounce-average is denoted by 
overhead bar. 
In Eq.~\eqref{eq:DriftKineticEqL11ziBounce} every term contains a factor $\nu_{z}'$ which can be divided away. 
The equation is a linear inhomogeneous equation for $f_{z1}$, where the inhomogeneous drive term is $\overline{C_{zi} \left[f_{Mz}, f_{i1}\right]}$ containing $f_{i1}$ and which is non-zero because of the $d n_i / d \psi$ drive in the main ion kinetic equation. 
Because of the $1/\nu$-regime of a stellarator in the absence of \new{$E_{\ast}$}, we expect that $f_{i1}$ scales like $1 / \nu_{z}'$ and it could thus be expected from Eq.~\eqref{eq:DriftKineticEqL11ziBounce} that also $f_{z1}$ scales as $1 / \nu_{z}'$ giving rise to the behavior in $\new{\tilde{L}_{11}^{zi}}$ we observe at low collisionality. 
Note that the $1/\nu$-part of $f_{i1}$ is even in $\vpa$, and since $\vpa$ parity is preserved by the field term of the linearized collision operator, then the even part of the $C_{zi}$ field term is required to couple this drive to the impurities. Hence, 
the $1/\nu_{z}'$-scaling of $\new{\tilde{L}_{11}^{zi}}$ will be missed in any numerical or analytic calculation in which the terms that are even in $\vpa$ are neglected in the field term of the collision operator. Momentum conservation, which is associated with the odd part of the field term, is not sufficient. For this reason, the momentum-corrected \dkes~results in Figs.~\ref{fig:CarbonTransportCoefficientsW7XZeff1p05}~(b) and \ref{fig:CarbonTransportCoefficientsW7XZeff2p0}~(b) obtain the wrong scaling with $\nu_{z}'$ and wrong sign at low collisionality. 
It is possible to test this hypothesis using a modified form of the impurity-ion Fokker-Planck\new{-Landau} collision operator in \sfincs, 
by selectively 
\new{turning}
off the field term for even Legendre modes. The results of these calculations for $Z_{\mathrm{eff}} = 1.05$ 
\new{at $E_{\ast} = 0$} 
are shown in Fig.~\ref{fig:L11ziComparison}, 
and clearly the $1 / \nu_{z}'$-behavior of $\new{\tilde{L}_{11}^{zi}}$ disappears when the even Legendre polynomials in the field term of $C_{zi}$ have been suppressed. 
Physically, if the bulk ions have a radial density gradient, then $f_{i1}$ carries an anisotropic pressure and will be rich in particles drifting outwards and poor in particles drifting inwards. 
The term $C_{zi} \left[f_{Mz}, f_{i1}\right]$ will try to create a similar anisotropy in $f_z$ and thus cause radial impurity transport. 

A negative $\new{\tilde{L}_{11}^{zz}}$ at all collisionalities is not surprising, 
since it merely tells us that a negative impurity density gradient will drive the impurities outwards. 
This is necessary and follows from the entropy law. 
Moreover, as shown in Appendix~\ref{sec:appendixB}, in the high-collisionality limit $\new{\tilde{L}_{11}^{zi}} = - Z \new{\tilde{L}_{11}^{zz}}$. 
\new{(We note that this is also approximately valid for the high-collisionality \sfincs~Fokker-Planck-Landau results at $E_{\ast} = -0.74$ 
in Fig.~\ref{fig:CarbonTransportCoefficientsW7XZeff2p0withEr20kV}, although \sfincs~retains $\Ev \times \Bv$-precession which is formally excluded in the usual drift ordering.)} 
This implies that the main ion density gradient will drive an impurity accumulation, 
which is stronger the higher the impurity charge. 
Furthermore, in accordance with the discussion for tokamaks in Ref.~\cite{helander02collisional}, in the Pfirsch-Schl\"uter regime the impurity transport is primarily driven by the bulk ion gradients. 
In the low-collisionality banana regime, however, $\new{\tilde{L}_{11}^{zi}}$ can be smaller in size than $\new{\tilde{L}_{11}^{zz}}$ and also be negative. This indicates that the bulk ion density gradient could mitigate an impurity accumulation in a hot reactor plasma. 
Note that since $\new{\tilde{L}_{11}^{zi}} = - Z \new{\tilde{L}_{11}^{zz}}$ in the Pfirsch-Schl\"uter limit, and because of our definition of the thermodynamic forces and the transport coefficients in Eqs.~\eqref{eq:ThermodynamicForces} and \eqref{eq:ImpFluxSFINCS}, 
the $\left(Ze/T_z\right) \left(d \Phi/d \psi\right)$ term in $\new{A_{z1}}$ is canceled by the $\left(e/T_i\right) \left(d \Phi/d \psi\right)$ term in $\new{A_{i1}}$ in Eq.~\eqref{eq:ImpFluxSFINCS}. 
The reason why the radial electric field has no effect on the impurity transport in this limit is 
that the transport is dominated by impurity-ion friction,  
and transport from friction 
is intrinsically ambipolar \cite{braun}. 
\new{However, it is important to emphasize that this is a result in the usual drift ordering where the $\Ev \times \Bv$-drift velocity is ordered $v_E \sim \rho_\ast v_i$ and $\Ev \times \Bv$-precession is formally excluded. Recall that \sfincs~(as well as other stellarator neoclassical calculations \cite{hirshmanDKES,vanRijDKES}) retains $\Ev \times \Bv$-precession in the drift-kinetic equation for $f_{a1}$, and it is thus possible that a finite $E_{\ast}$ yields a different Pfirsch-Schl\"uter limit than $E_{\ast} = 0$ in \sfincs~calculations. 
As noted in Ref.~\cite{beidlerNF11}, poloidal $\Ev \times \Bv$-precession is traditionally ignored in the Pfirsch-Schl\"uter regime but can become relevant when the product of the radial electric field and collisionality is sufficiently large.} 

The $\new{\tilde{L}_{12}^{z}} - 3 \left(\new{\tilde{L}_{11}^{zz}} + \new{\tilde{L}_{11}^{zi}}\right)/2$ coefficient illustrated in 
Figs.~\ref{fig:CarbonTransportCoefficientsW7XZeff1p05}-\NewSecond{\ref{fig:CarbonTransportCoefficientsW7XZeff2p0}}~(d)
is the temperature gradient coefficient (recall that we assume $T_i = T_z$), i.e. the coefficient in front of $d \ln T / d \psi$ when substituting Eq.~\eqref{eq:ThermodynamicForces} into Eq.~\eqref{eq:ImpFluxSFINCS}. 
This coefficient corresponds to $- \left(D_2^{zz} + D_2^{zi}\right)$ in Eq.~\eqref{eq:NeoPartFlux}, adjusted with a normalization factor. 
A negative value indicates temperature screening which is found 
in the low-collisionality regime. 
Figure~\ref{fig:CarbonTransportCoefficientsW7XZeff2p0}~(d) shows that 
in the high-collisionality regime \sfincs~finds that the temperature gradient drives the impurities inwards, with a strength increasing 
with collision frequency 
when $Z_{\mathrm{eff}} = 2.0$ \new{and $E_{\ast} = 0$}, 
which is in agreement with the analytical Pfirsch-Schl\"uter asymptotes \NewSecond{(see Appendix~\ref{sec:appendixB})}. 
For $Z_{\mathrm{eff}} = 1.05$ \new{and $E_{\ast} = 0$} the \sfincs~Fokker-Planck\new{-Landau} calculations show a temperature screening also in the high-collisionality regime, 
as illustrated in Fig.~\ref{fig:CarbonTransportCoefficientsW7XZeff1p05}~(d). 
Our \sfincs~results are again in agreement with the analytical Pfirsch-Schl\"uter asymptotes, and it is intriguing to see that whether we find a temperature screening in the Pfirsch-Schl\"uter limit or not can, in fact, depend on the impurity content. 
It should be emphasized that these extremely high collisionalities are irrelevant in practice for at least two reasons. 
Firstly, the observed transport is usually turbulent in plasmas that are cold enough to be in the Pfirsch-Schl\"uter regime; 
and secondly, the neoclassical transport is sensitive to the radial electric field. 

Since $\new{\tilde{L}_{11}^{zz}}$, $\new{\tilde{L}_{11}^{zi}}$ and $\new{\tilde{L}_{12}^{z}} - 3 \left(\new{\tilde{L}_{11}^{zz}} + \new{\tilde{L}_{11}^{zi}}\right)/2$ (i.e. the impurity density gradient, the ion density gradient and the temperature gradient coefficients respectively) are all negative in the low-collisionality regime \new{at $E_{\ast} = 0$}, 
\new{one might think} 
that both the main ion and impurity pressure gradients would be beneficial for avoiding neoclassical impurity accumulation in a hot reactor-like plasma if the profiles are peaked. 
However it should be remembered that the radial impurity transport will typically heavily depend on the ambipolar radial electric field which builds up to balance the particle fluxes and to yield a vanishing radial net current. 


\NewSecond{
\subsection*{Simulation results at finite $E_r$}\label{subsec:SimulationResultsFiniteEr}
To examine the role of the radial electric field, we perform 
\sfincs~calculations with full linearized Fokker-Planck-Landau collisions
and $Z_{\mathrm{eff}} = 2.0$ at $E_{\ast} = -0.74$. 
This value of $E_{\ast}$ corresponds to $E_r = -20~\mathrm{kV/m}$ if the temperature is $T = 1~\mathrm{keV}$. 
The results are illustrated in Fig.~\ref{fig:CarbonTransportCoefficientsW7XZeff2p0withEr20kV}   
and when comparing the \sfincs~curves to the \dkes~curves it should be recalled that the simulations employ different effective particle trajectories, which matters when $E_{\ast} \neq 0$. 
}
\NewThird{
To examine if the difference between the \sfincs~results and the \dkes~results is mainly a consequence of the different collision operators or the different effective particle trajectories which are employed in the two tools, we have also included results from \sfincs~calculations using ``\dkes~particle trajectories''. As shown in Fig.~\ref{fig:CarbonTransportCoefficientsW7XZeff2p0withEr20kV} the effect of the different models for the particle trajectories is small for this particular case. However, note that as $E_{\ast}$ approaches unity a difference can in general  
be expected \cite{landremanSFINCS}. 
}
\NewSecond{
In contrast to the results for $E_{\ast} = 0$, 
at $E_{\ast} = -0.74$ 
there is no sign change in the \sfincs~Fokker-Planck-Landau curves for $\tilde{L}_{11}^{zi}$ at low collisionality, as seen in Fig.~\ref{fig:CarbonTransportCoefficientsW7XZeff2p0withEr20kV}~(b), but there is still a difference to the \dkes~+ momentum correction results. 
Moreover, the $1/\nu_{z}'$-transport at low collisionality and $\nu_{z}'$-proportionality at high collisionality seen in Figs.~\ref{fig:CarbonTransportCoefficientsW7XZeff1p05} and \ref{fig:CarbonTransportCoefficientsW7XZeff2p0} disappear for the finite $E_{\ast}$. 
The temperature coefficient 
shown in Fig.~\ref{fig:CarbonTransportCoefficientsW7XZeff2p0withEr20kV}~(d) indicates a temperature screening at all plotted collisionalities, but the effect is weak at low collisionality.
In the low-collisionality regime 
we also find a negative impurity density gradient coefficient, 
but a small positive ion density gradient coefficient. 
The magnitude of $\tilde{L}_{11}^{zi}$ is significantly smaller than the magnitude of $\tilde{L}_{11}^{zz}$, which implies that if we substitute a negative $E_r$ into Eqs.~\eqref{eq:ThermodynamicForces} and \eqref{eq:ImpFluxSFINCS} the electric field will cause impurity accumulation as expected. 
\new{
\begin{figure}[!ht]
\begin{center}
\includegraphics[width=1.0\textwidth]{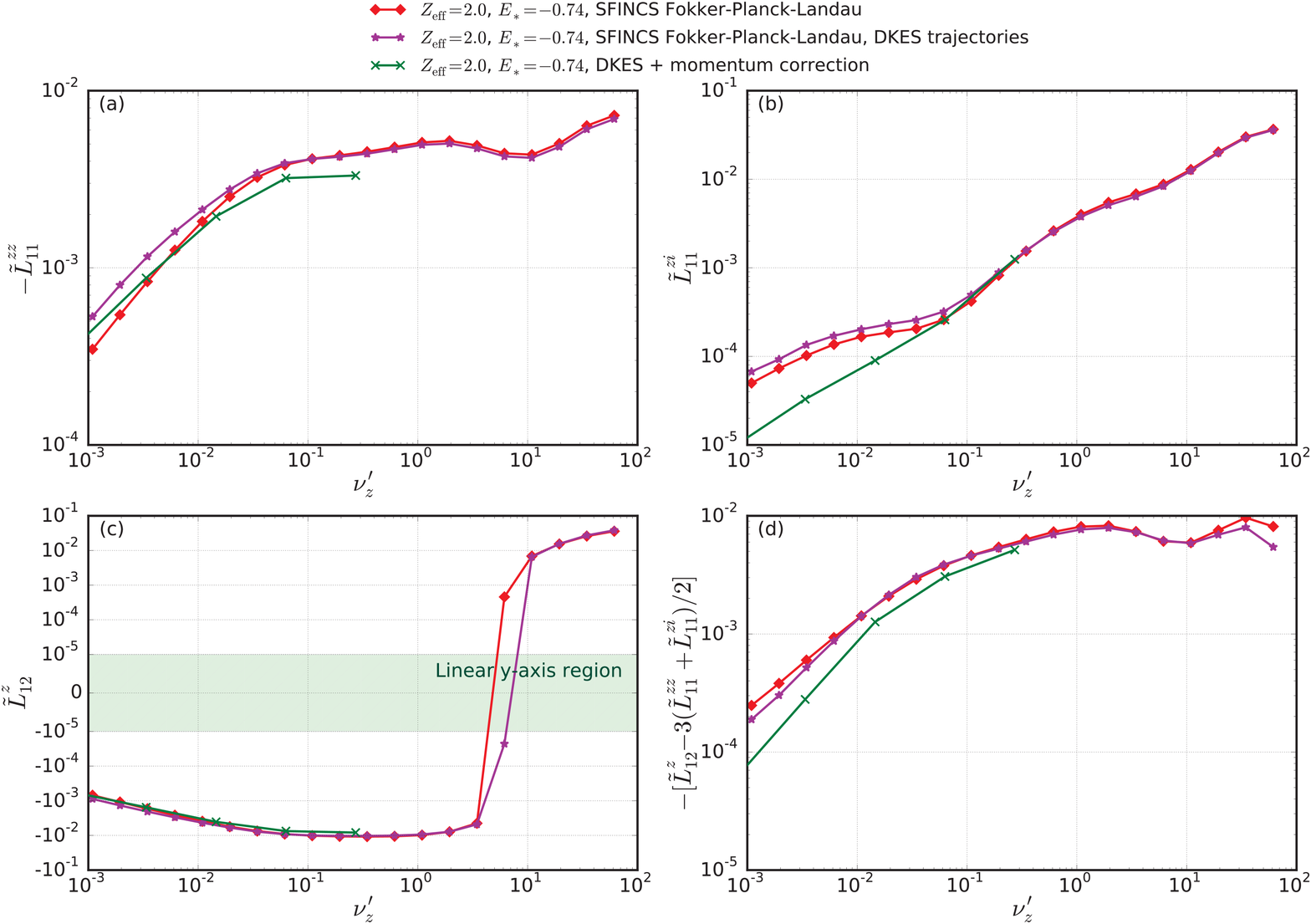}
\end{center}
\caption{\new{
Carbon ($Z = 6$) transport coefficients $\new{\tilde{L}_{11}^{zz}}$ (a), $\new{\tilde{L}_{11}^{zi}}$ (b), $\new{\tilde{L}_{12}^{z}}$ (c) and the temperature gradient coefficient $\new{\tilde{L}_{12}^{z}} - 3 \left(\new{\tilde{L}_{11}^{zz}} + \new{\tilde{L}_{11}^{zi}}\right)/2$ (d) as functions of normalized collision frequency $\nu_{z}'$ for a W7-X geometry at $E_{\ast} = -0.74$ 
and with $Z_{\mathrm{eff}} = 2.0$. 
\NewSecond{\sfincs~computations with the Fokker-Planck-Landau collision operator (\textcolor{Red}{\fulldiamondline}) 
are compared to \dkes~computations with momentum correction applied afterwards 
(\textcolor{Green}{\crossline}).
}
\NewThird{Also shown are \sfincs~computations using a different model for the particle trajectories, referred to as 
``\dkes~particle trajectories'' and described in Ref.~\cite{landremanSFINCS} (\textcolor{NewMagenta}{\starline}).}
Note the double-logarithmic scale in (c).} 
}
\label{fig:CarbonTransportCoefficientsW7XZeff2p0withEr20kV}
\end{figure}
}
}


\new{As a final remark of this section, we note that we can also use \sfincs~to calculate the main ion transport coefficients in Eq.~\eqref{eq:TransportMatrix}. 
We find that $L_{11}^{iz} = L_{11}^{zi}$ \NewSecond{(both for finite and vanishing $E_{\ast}$)}, and thus the Onsager symmetry described in the beginning of the section is fulfilled.
}\\


\section{Impurity density peaking and bootstrap current}\label{sec:impurity}
In a 3D device, a bootstrap current arises due to similar reasons as in a tokamak, but the size of it is typically substantially smaller than the Ohmic current in a tokamak \cite{fusionphysics12,helanderStellarator,helanderBootstrap}. 
The bootstrap current is a consequence of the trapped particle orbits, and is generally larger at low collisionality than at high collisionality 
\cite{helander02collisional,landremanBootstrap}. 
Since the bootstrap current adds pressure-dependence to the magnetic equilibrium, 
it can be desirable in stellarators to minimize the bootstrap current so the magnetic field remains optimized over a range of plasma pressure. 
W7-X has been optimized for a small bootstrap current. 
A net toroidal current changes the value of $\iota$ at the boundary, which can be detrimental for proper island divertor operation \cite{geigerBootstrap1,geigerBootstrap2,loreBootstrap}. 
It is consequently important to be able to make realistic predictions of the bootstrap current when designing a stellarator. 

\begin{figure}[!ht]
\begin{center}
\includegraphics[width=0.8\textwidth]{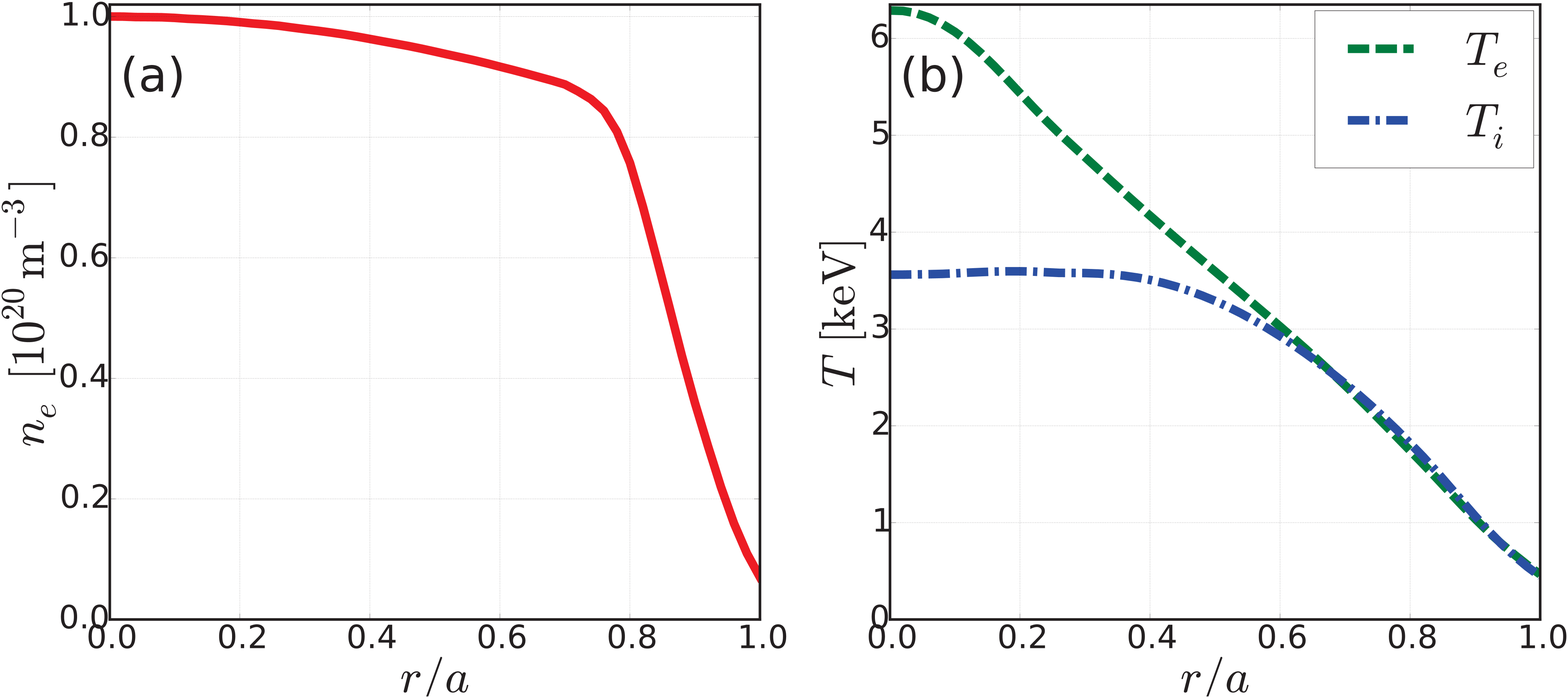}
\end{center}
\caption{
The plasma profiles for a W7-X standard magnetic configuration corresponding to central electron cyclotron resonance heating profile and an average $\beta$ of $2.9$~\%: (a)~assumed electron density (\textcolor{Red}{$\full$}); 
(b)~predicted electron (\textcolor{Green}{$\broken$}) and ion temperatures (\textcolor{Blue}{$\longchain$}). 
}
\label{fig:RadialProfiles}
\end{figure}

In this section we use \sfincs~to investigate how the presence of impurities affects the bootstrap current in a non-axisymmetric plasma, 
and also how the neoclassical impurity dynamics are affected by the impurity content through the plasma effective charge. 
Again we study a hydrogen plasma with a single carbon impurity species present (this time the electrons are included in our simulations), and use 
the W7-X standard 
magnetic configuration. 
In contrast to Sec.~\ref{sec:TransportCoefficients}, the configuration here corresponds to a central electron cyclotron resonance heating profile and an average $\beta$ of $2.9$~\%, 
$\beta \equiv 8 \pi n T / B^2$ being the ratio of plasma pressure to magnetic pressure. 
The 
\NewSecond{temperature} 
profiles have been calculated according to the procedure in Ref.~\cite{TurkinPoP}, \new{in which the electron density profile is assumed on the basis of other experiments}. 
The density and temperature profiles are shown in Fig.~\ref{fig:RadialProfiles}. 

To study the radial impurity transport, we calculate the neoclassical zero-flux impurity density gradient (also referred to as the impurity peaking factor) 
defined as $a / L_{nz}$ for which the flux-surface-averaged impurity flux vanishes, 
$
\left\langle \Gamma_z \cdot \na \psi\right\rangle 
= 0$. 
This happens when the convective part of the transport is balanced by the diffusive part, corresponding to $V_z$ and $D_{1}^{zz}$ in Eq.~\eqref{eq:ImpurityFluxEquation}. 
$L_{nz} = - \left[d \ln n_z  / d r\right]^{-1}$ is the impurity density gradient scale length. 
(Note that $L_{T} = - \left[d \ln T  / d r\right]^{-1}$ is kept fixed when calculating the zero-flux impurity density gradient.) 
To calculate the impurity peaking factor we must also find the ambipolar radial electric field simultaneously, 
i.e. the $E_r$ for which the radial net current vanishes, $\displaystyle \sum_b Z_b e \left\langle \Gamma_b \cdot \na \psi\right\rangle = 0$. 
As earlier mentioned, in a typical ion root scenario (negative $E_r$) where the impurity concentration is small compared to the main species, 
the ambipolar electric field arises to bring the main ion particle transport down to the electron level, and 
so the ambipolar electric field is approximately the field 
for which $\displaystyle \left\langle \Gamma_i \cdot \na \psi\right\rangle \simeq 0$.
For a single impurity species plasma, with the impurity species in trace contents, it is always possible to find values of $d n_z / d \psi$ and $E_r$ such that the radial impurity flux and the radial current vanish simultaneously. 
We note however, that if $Z_{\mathrm{eff}} = Z$ (i.e. a plasma consisting of only electrons and the impurity species) ambipolarity requires the impurity flux to balance the outward electron flux, and it is not possible to find a neoclassical impurity peaking factor. Thus, there must be a critical value of $Z_{\mathrm{eff}}$ between 1 and $Z$ above which the radial impurity flux and the radial current cannot vanish simultaneously. (This is the reason why we 
later in this section 
are not able to present the impurity peaking factor at large values of $Z_{\mathrm{eff}}$ for 
one of the cases, 
i.e. no solution exists.)

\begin{figure}[!ht]
\begin{center}
\includegraphics[width=1.0\textwidth]{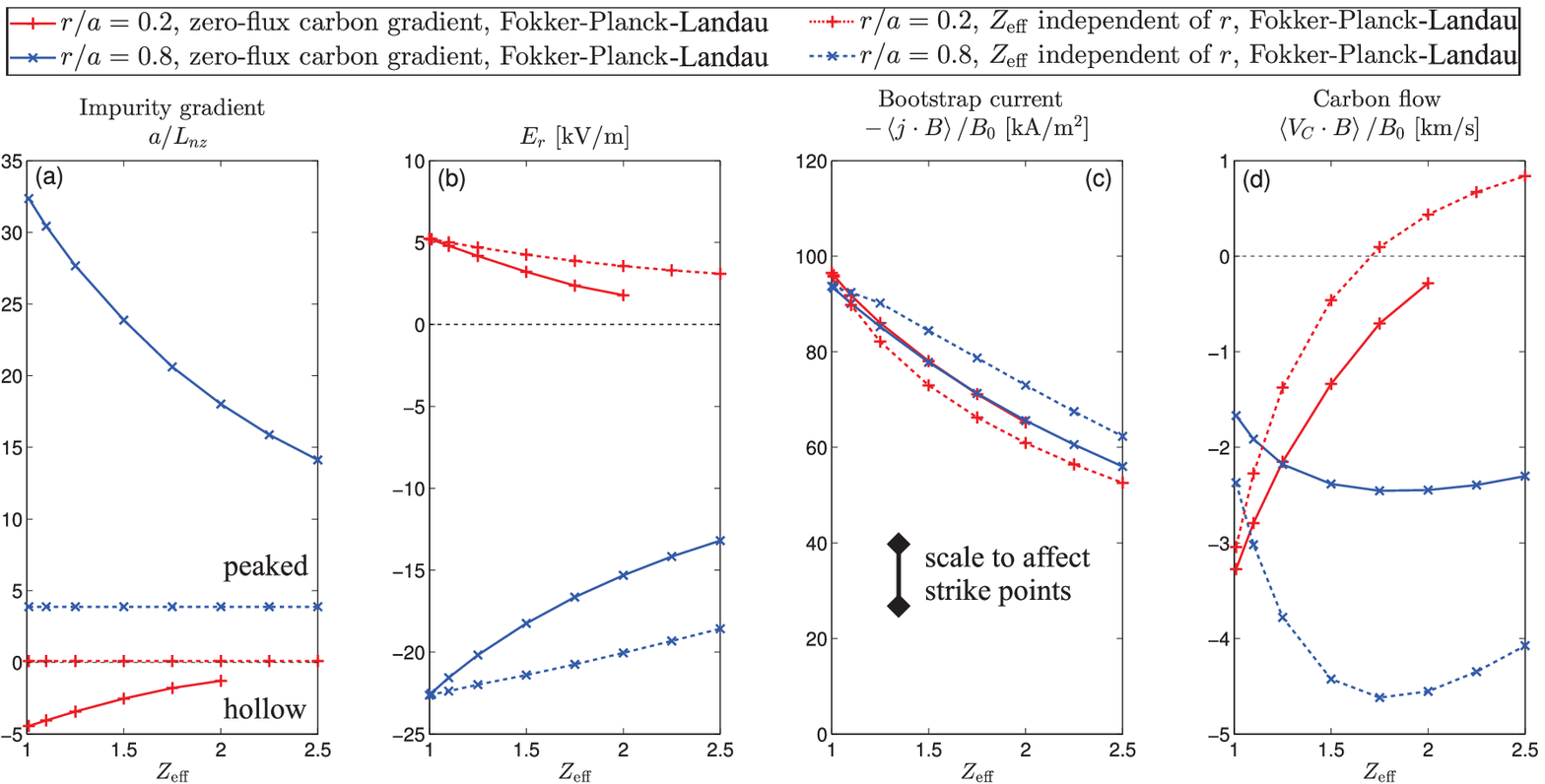}
\end{center}
\caption{Carbon ($Z = 6$) density gradient (a), ambipolar radial electric field (b), bootstrap current density (c), and carbon flow (d) as functions of the plasma effective charge for a W7-X geometry. 
\sfincs~computations 
with Fokker-Planck-Landau collisions 
at two different radii are compared, using two different models for the ion density gradients, 
one when the carbon peaking factor is found (solid lines labeled ``zero-flux carbon gradient'') and another when $Z_{\mathrm{eff}}$ is kept independent of the radial location (dashed lines labeled ``$Z_{\mathrm{eff}}$ independent of $r$''): 
$r/a = 0.2$ 
(\textcolor{Red}{${\bm +}$}), 
$r/a = 0.8$ 
(\textcolor{Blue}{${\bm \times}$}). 
}
\label{fig:W7XjbsVsZeff}
\end{figure}

We analyze two radial locations, $r/a = 0.2$ and $r/a = 0.8$. 
At $r/a = 0.2$ the 
parameters are 
$n_e = 0.991 \times 10^{20}~\mathrm{m}^{-3}$, $T_e = 5.44~\mathrm{keV}$, $T_i = 3.60~\mathrm{keV}$, $\displaystyle d n_e / d r = - 0.209\times 10^{20}~\mathrm{m}^{-4}$, $\displaystyle d T_e / d r = -13.5~\mathrm{keV}/\mathrm{m}$ and $\displaystyle d T_i / d r = -0.269~\mathrm{keV}/\mathrm{m}$. 
At $r/a = 0.8$ they are $n_e = 0.758 \times 10^{20}~\mathrm{m}^{-3}$, $T_e = 1.74~\mathrm{keV}$, $T_i = 1.82~\mathrm{keV}$, $\displaystyle d n_e / d r = -6.93 \times 10^{20}~\mathrm{m}^{-4}$, $\displaystyle d T_e / d r = -13.1~\mathrm{keV}/\mathrm{m}$ and $\displaystyle d T_i / d r = -13.1~\mathrm{keV}/\mathrm{m}$. 
$Z_{\mathrm{eff}}$ is varied by varying the impurity density (and main ion density accordingly to fulfill quasi-neutrality at fixed $n_e$), and this implies that the normalized collision frequency, defined in Eq.~\eqref{eq:nuPrimeZ}, satisfies $\nu_{z}' \lesssim 0.25$ for $r/a = 0.2$ and $\nu_{z}' \lesssim 0.70$ for $r/a = 0.8$ respectively. 
It is difficult to anticipate on what time scales the impurity species will reach a steady-state 
(and its corresponding zero-flux impurity density gradient), and also how the main species density gradients will adapt to keep radial quasi-neutrality. 
We therefore carry out two different scans in $Z_{\mathrm{eff}}$. In the first scan we find the impurity peaking factor, and the main ion density gradient is modified along with the impurity density gradient to preserve radial quasi-neutrality, while the electron density gradient is kept fixed. 
(The corresponding curves are solid and labeled ``zero-flux carbon gradient'' in Fig.~\ref{fig:W7XjbsVsZeff}.) 
In the second scan we instead keep the radial density gradient scale lengths fixed and equal, $L_{ne} = L_{ni} = L_{nz}$, which is equivalent to the condition $\displaystyle d Z_{\mathrm{eff}} / d r = 0$. (The corresponding curves are dashed and labeled ``$Z_{\mathrm{eff}}$ independent of $r$'' in Fig.~\ref{fig:W7XjbsVsZeff}.) 

Figure~\ref{fig:W7XjbsVsZeff} shows our \sfincs~results for the carbon density gradient, ambipolar radial electric field, bootstrap current density, and carbon flow as functions of $Z_{\mathrm{eff}}$ for the W7-X geometry. The scans are performed with 
full linearized Fokker-Planck-Landau collisions. 
%
Comparing the approach when we find the carbon peaking factor to the approach when we keep the density gradient scale lengths fixed, we see that qualitatively a similar ambipolar electric field and bootstrap current density are obtained (compare the solid and dashed curves in Figs.~\ref{fig:W7XjbsVsZeff}~(b) and (c)). Unsurprisingly the deviation between the two approaches increases as $Z_{\mathrm{eff}}$ becomes larger, mainly because the difference in main ion density gradient also becomes larger. A similar argument, but regarding the impurity density gradient, is likely the reason why there is a difference in parallel carbon flow between the two approaches (see Fig.~\ref{fig:W7XjbsVsZeff}~(d)). 
Note that particularly the core carbon flow (both the direction and the magnitude) at $r/a = 0.2$ is sensitive to $Z_{\mathrm{eff}}$. Thus the flow could make a sensitive diagnostic test of neoclassical physics. 

\begin{figure}[!ht]
\begin{center}
\includegraphics[width=0.8\textwidth]{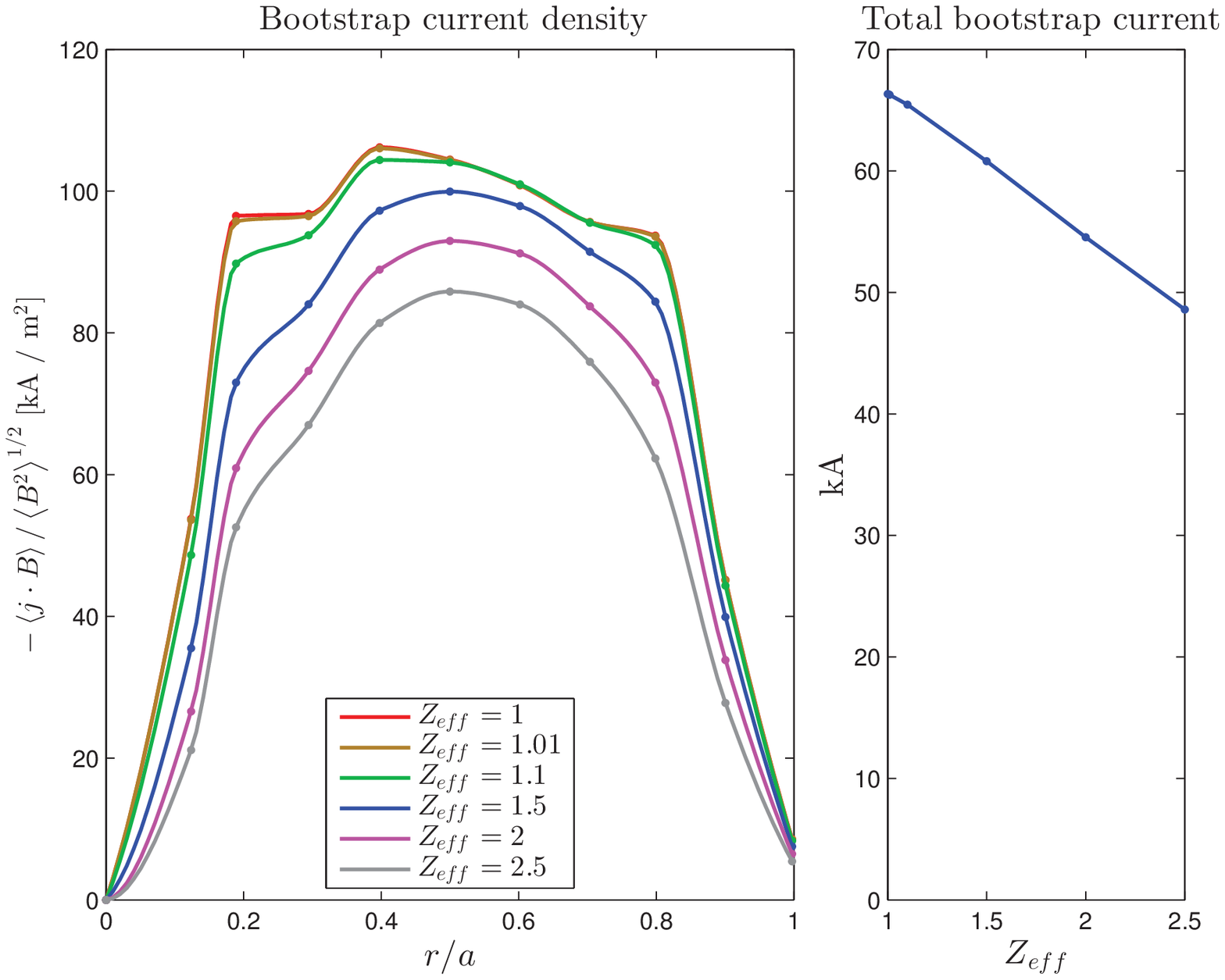}
\end{center}
\caption{
Radial profiles of the bootstrap current density from \sfincs~calculations with the full linearized Fokker-Planck\new{-Landau} collision operator at different $Z_{\mathrm{eff}}$ (keeping $Z_{\mathrm{eff}}$ constant throughout the radial domain) and the integrated total bootstrap current as a function of $Z_{\mathrm{eff}}$.
}
\label{fig:W7XjbsVsZeffconstZeffProfiles}
\end{figure}

Irrespective of which 
model for the density gradients we use, some main features are clearly observed. 
Firstly, at $r/a = 0.2$ the plasma is in the electron root regime and at $r/a = 0.8$ in the ion root regime, since the ambipolar radial electric field has the opposite sign as shown in  Fig.~\ref{fig:W7XjbsVsZeff}~(b). 
Furthermore, in accordance with what was found in Ref.~\cite{mollenVarenna} 
the ambipolar electric field is reduced as the impurity content is increased. 
This also implies that the impurity profile, which is hollow at $r/a = 0.2$ and peaked at $r/a = 0.8$, is flattened as illustrated in Fig.~\ref{fig:W7XjbsVsZeff}~(a). 
The bootstrap current density is also significantly reduced with increased impurity content, although no sign change is observed (see Fig.~\ref{fig:W7XjbsVsZeff}~(c)). 
This reduction is not surprising, since the electron-ion friction increases with $Z_{\mathrm{eff}}$.  
The results show that changes in $Z_{\mathrm{eff}}$ on the order of $\Delta Z_{\mathrm{eff}} \sim 1$ can lead to changes in the 
bootstrap current density on the order of $\Delta j_{\mathrm{bs}}\gtrsim 20 \mathrm{kA/m^2}$.
If such changes occurred across the entire plasma cross-section, 
the total current could change by $\Delta I_{\mathrm{bs}} \gtrsim  10~\mathrm{kA}$. 
This is illustrated in Fig.~\ref{fig:W7XjbsVsZeffconstZeffProfiles}, where the integrated total bootstrap current is shown as a function of $Z_{\mathrm{eff}}$ 
when $Z_{\mathrm{eff}}$ is kept constant throughout the radial domain. 
A change in the plasma current of $10~\mathrm{kA}$ 
could modify 
the value of the boundary-$\iota$
enough to cause measurable changes in the island divertor strike point locations \cite{geigerBootstrap1,geigerBootstrap2,loreBootstrap}, 
indicating that the full ion composition should be considered when performing bootstrap current calculations. 


\section{Discussion and conclusions}
\label{sec:conclusions}
In this work we have used a continuum drift-kinetic solver, the \sfincs~code, to calculate neoclassical impurity transport coefficients (defined by Eqs.~\eqref{eq:ThermodynamicForces} and \eqref{eq:ImpFluxSFINCS})
in a non-axisymmetric magnetic equilibrium. 
Particularly, we studied carbon transport close to the plasma edge in a W7-X hydrogen plasma 
for two different levels of the impurity content corresponding to $Z_{\mathrm{eff}} = 1.05$ and $Z_{\mathrm{eff}} = 2.0$ respectively, 
at vanishing radial electric field. 
\new{For $Z_{\mathrm{eff}} = 2.0$ we also studied the carbon transport at $E_{\ast} = -0.74$ (corresponding to $E_r = -20~\mathrm{kV/m}$ if the temperature is $T = 1~\mathrm{keV}$).}

\NewSecond{We compare \sfincs~computations with full linearized Fokker-Planck-Landau collisions to computations with the \dkes~code (which employs pitch-angle scattering) where momentum correction is applied afterwards. 
We find that the impurity density gradient and temperature gradient coefficients are well reproduced by the numerical model with pitch-angle scattering and momentum correction, but to correctly determine the inter-species ion density gradient coefficient it is sometimes not sufficient to merely account for momentum conservation.}

The impurity transport coefficients show trends of $1/\nu_{z}'$-transport at low collisionality and are proportional to $\nu_{z}'$ at high collisionality \new{if $E_{\ast} = 0$}. 
From earlier work $1/\nu_{z}'$-transport is not expected for the ion density gradient coefficient. 
We show that, by suppressing the even Legendre polynomials in the field term of the impurity-ion collision operator in \sfincs~the $1/\nu_{z}'$-transport for this coefficient disappears. 
Earlier work has often approximated the field-particle part of the collision operator by a momentum-conserving 
term, 
which has the wrong $\vpa$ parity to couple the (even) $1/\nu_{z}'$-part of $f_{i1}$ to the impurities 
and this is likely the reason why $1/\nu_{z}'$-transport for this cross-species transport coefficient has not been observed. 
\NewSecond{If we introduce a finite $E_{\ast}$ in our calculations, we instead find trends of $\sqrt{\nu_{z}'}$-transport at low collisionality.}
Not surprisingly, we find that the impurity density gradient coefficient is negative at all collisionalities which merely implies that 
a standard negative gradient drives the impurities outwards. 
At high collisionality however, the impurity transport is dominated by the bulk ion density gradient (by a factor $Z$) which drives the impurities inwards. 
At low collisionality we find that all transport coefficients are negative \new{when $E_{\ast} = 0$}, indicating an impurity screening. 
This could be beneficial from a reactor point of view, since the hot (almost collisionless) core could avoid impurity accumulation. 
\new{However, when we introduce a negative ambipolar radial electric field (ion root regime) it drives impurity accumulation as expected.}
Interestingly we find a temperature screening in the low-collisionality regime which persists up to a relatively high collisionality for $Z_{\mathrm{eff}} = 2.0$, 
and is maintained at all collisionalities for $Z_{\mathrm{eff}} = 1.05$. 
\new{This is also the case for the calculations with a finite $E_{\ast}$.} 
In the high collisionality limit, an analytic prediction is available for the transport coefficients \new{at $E_{\ast} = 0$}, and the \sfincs~calculations conform well with these predictions. 

Moreover, we have used \sfincs~to investigate how the impurity content affects the neoclassical impurity dynamics and the bootstrap current in a W7-X plasma. 
We find that an increased impurity content, implying a higher plasma effective charge, tends to flatten the impurity profile 
(determined from the condition of zero impurity particle flux) 
both close to the core ($r/a = 0.2$) 
where it is hollow and close to the edge ($r/a = 0.8$) where it is peaked. 
This trend is attributed to the reduction 
of the ambipolar radial electric field with increasing $Z_{\mathrm{eff}}$. 
The bootstrap current is also reduced with increasing impurity content, which is expected since the electron-ion friction increases with $Z_{\mathrm{eff}}$. 
Importantly, we find that the change in bootstrap current can be larger than $10~\mathrm{kA}$ for a change in $Z_{\mathrm{eff}}$ of $\ord \left(1\right)$. 
A change of this size could be large enough to cause a deviation in the divertor strike point locations.  
This emphasizes 
the importance of performing bootstrap current calculations with a realistic ion composition.


\section*{Acknowledgments}
The authors are grateful to J.~Geiger for providing the W7-X equilibrium data, to Y.~Turkin for providing simulated W7-X plasma profiles, 
and to I.~Pusztai, T.~F\"ul\"op, C.~D.~Beidler and H.~Maa{\ss}berg for input on the work. 
ML was supported by the U.S. Department of Energy, Office of Science, Office of Fusion Energy Science, under Award Numbers DEFG0293ER54197 and DEFC0208ER54964. 
For some of the computations, the Max Planck supercomputer at the Garching Computing Center RZG was used. 
Some computations were also performed on the Edison system at the National Energy Research Scientific Computing Center, a DOE Office of Science User Facility supported by the
Office of Science of the U.S. Department of Energy under Contract No. DE-AC02-05CH11231.


\section*{Appendix}
\renewcommand{\theequation}{A\arabic{equation}}
\renewcommand{\theHequation}{A\arabic{equation}}
\setcounter{equation}{0}
\setcounter{section}{0}
\renewcommand\thesection{\Alph{section}}
\renewcommand\theHsection{\Alph{section}}

\new{
\section{Collision operator in SFINCS}\label{sec:appendixA}
The total collision operator for species $a$ is a sum of linearized collision operators with each species: 
$ \displaystyle
C_a\left[f_a\right] \equiv \sum_b C_{ab}^{l}\left[f_a, f_b\right],
$
where 
$ \displaystyle
C_{ab}^{l}\left[f_a, f_b\right] \equiv C_{ab}\left[f_{a1}, f_{Mb}\right] + C_{ab}\left[f_{Ma}, f_{b1}\right]
$
and $C_{ab}\left[f_{a}, f_{b}\right]$ is the Fokker-Planck\new{-Landau} collision operator. 
$C_{ab}\left[f_{a1}, f_{Mb}\right]$ 
is referred to as the test particle part 
and $C_{ab}\left[f_{Ma}, f_{b1}\right]$ is the field particle part \cite{helander02collisional,RosenbluthPotentials,landremanComput}.
The linearized collision operator may be written as 
$ \displaystyle
C_{ab}^{l} = C_{ab}^{L} + C_{ab}^{E} + C_{ab}^{F},
$
where $C_{ab}^{L} + C_{ab}^{E}$ together represent the test particle part and $C_{ab}^{F}$ the field particle part.
The Lorentz part is 
$ \displaystyle
C_{ab}^{L}  = \nu_{Dab}\left(v\right) \, \mathcal{L} \left[f_{a}\right]
$
with $\displaystyle \nu_{Dab}\left(v\right) \equiv \hat{\nu}_{ab} \frac{\phi\left(x_b\right) - \Psi\left(x_b\right)}{x_a^3}$ being the deflection frequency and 
$\displaystyle \mathcal{L} \equiv \frac{1}{2} \frac{\p }{\p \xi} \left(1 - \xi^2\right) \frac{\p }{\p \xi}$ 
the Lorentz operator. 
In this context $\displaystyle \phi\left(x\right) = \frac{2}{\sqrt{\pi}} \int_0^x \exp \left(-y^2\right) dy$ is the error function, $\displaystyle \Psi\left(x\right) = \frac{\phi\left(x\right) - x \phi'\left(x\right)}{2 x^2}$ is the Chandrasekhar function, 
$\hat{\nu}_{ab} = 2^{1/2} \pi n_b e_a^2 e_b^2 \ln \Lambda/\left(m_a^{1/2} T_a^{3/2}\right)$ and $x_a = v / v_a$ with $v_a =\sqrt{2T_a/m_a}$ being the thermal speed of the species $a$. 
The energy scattering contribution 
is 
\begin{equation}
C_{ab}^{E}  = \nu_{\| ab} \left[\frac{v^2}{2} \frac{\p^2 f_{a1}}{\p v^2} - x_b^2 \left(1 - \frac{m_a}{m_b}\right) v \frac{\p f_{a1}}{\p v}\right] + 
\nu_{Dab} v \frac{\p f_{a1}}{\p v} + 4 \pi \frac{v_a^3}{n_b} \hat{\nu}_{ab} \frac{m_a}{m_b} f_{Mb} f_{a1},
\label{eq:LinearizedCollisionOperatorEnergyScatteringPart}
\end{equation}
where
$ \displaystyle
\nu_{\| ab} = 2  \hat{\nu}_{ab} \frac{\Psi\left(x_b\right)}{x_a^3}.
$
We can write the field term as
\begin{equation}
C_{ab}^{F}  = C_{ab}^{H}  + C_{ab}^{G}  + C_{ab}^{D},
\label{eq:FieldTermParts}
\end{equation}
where 
\begin{equation}
C_{ab}^{H} = \frac{\hat{\nu}_{ab}}{n_b} f_{Ma} \left[- 2 v v_a \left(1 - \frac{m_a}{m_b}\right) \frac{\p H_{b 1}}{\p v} - 2 v_a H_{b 1}\right],
\label{eq:CHpart}
\end{equation}
\begin{equation}
C_{ab}^{G} = \frac{\hat{\nu}_{ab}}{n_b} f_{Ma} \frac{2 v^2}{v_a} \frac{\p^2 G_{b 1}}{\p v^2}
\label{eq:CGpart}
\end{equation}
and
\begin{equation}
C_{ab}^{D} = \frac{\hat{\nu}_{ab}}{n_b} v_a^3 f_{Ma} \, 4 \pi \frac{m_a}{m_b} f_{b1}.
\label{eq:CDpart}
\end{equation}
The functions $G_{b 1}$ and $H_{b 1}$ are the perturbed Rosenbluth potentials \cite{RosenbluthPotentials} defined by 
$ \displaystyle
\na_v^2 H_{b 1} = - 4 \pi f_{b1}
$
and
$ \displaystyle
\na_v^2 G_{b 1} = 2 H_{b 1}.
$
}

\new{
The speed discretization in \sfincs~is based on a spectral collocation scheme described in \cite{landremanComput}: 
a function $f \left(x\right)$ is 
stored at grid points $x_j$ which are the zeros of a polynomial, where the polynomial is taken from the set 
$M_n^k \left(x\right)$ (with $n \geq 0$) obeying the orthogonality relation 
\begin{equation}
\int_0^{\infty} d x \, x^k \exp \left(- x^2\right) M_n^k \left(x\right) M_m^k \left(x\right) = \delta_{n,m} A_n^k.
\label{eq:OrthogonalityRelation}
\end{equation}
Here $\delta_{n,m}$ is the Kronecker delta, $A_n^k$ represents some normalization and 
$k$ is any number greater than $-1$, but from experience the choice $k = 0$ is typically good. 
Note that $x$ here denotes the speed normalized to thermal speed, e.g. the distribution function for species $a$ is $f_a \left(x_a\right)$. 
We can alternatively represent $f$ in a modal discretization by the vector of numbers $F_n^k$ in 
\begin{equation}
f \left(x\right) = \sum_n F_n^k M_n^k \left(x\right) \exp \left(- x^2\right),
\label{eq:fModalExpansion}
\end{equation}
where
\begin{equation}
F_m^k = \frac{1}{A_m^k} \int_0^{\infty} d x \, x^k M_m^k \left(x\right) f \left(x\right).
\label{eq:Fnk}
\end{equation} 
In terms of the Gaussian integration weights $w_j$ associated with the grid $x_j$ satisfying $\displaystyle \int_0^\infty dx\;y(x) \approx \sum_j w_j y(x_j)$, there is thus a linear transformation $Y$ from the collocation to the modal discretization, with $Y_{m,n} = w_n x_n^k M_m^k(x_n)/A_m^k$.
}

\new{
At each speed grid point, the pitch-angle dependence of the distribution function is decomposed in Legendre polynomial modes:
\begin{equation}
f_{a1}\left(x_a,\xi\right)=\sum_l P_l\left(\xi\right)f_{a1, l}\left(x_a\right).
\label{eq:fb1LegendreModes}
\end{equation}
In the Legendre modal representation, $C_{ab}^L$ becomes diagonal. As in Ref.~\cite{landremanComput}, $C_{ab}^E$ can be represented using the pseudospectral differentiation matrix associated with the polynomials $M_n^k(x)$, and $C_{ab}^D$ can be represented using the interpolation matrix associated with the $M_n^k(x)$. In evaluating $C_{ab}^H$ and $C_{ab}^G$, however, we depart from the method in \cite{landremanComput}. First an expansion analogous to Eq.~\eqref{eq:fb1LegendreModes} is made for
the perturbed potentials in terms of their Legendre modes $H_{b1, l} \left(x_b\right)$ and $G_{b1, l} \left(x_b\right)$, 
and 
we 
find 
\begin{equation}
\frac{\p}{\p x_b} x_b^2 \frac{\p H_{b 1, l}}{\p x_b} - l \left(l + 1\right) H_{b 1, l} =  - 4 \pi v^2 f_{b1, l},
\label{eq:PoissonEqHb1}
\end{equation}
\begin{equation}
\frac{\p}{\p x_b} x_b^2 \frac{\p G_{b 1, l}}{\p x_b} - l \left(l + 1\right) G_{b 1, l} =  2  v^2 H_{b 1, l}.
\label{eq:PoissonEqGb1}
\end{equation}
Solving Eqs.~\eqref{eq:PoissonEqHb1}-\eqref{eq:PoissonEqGb1} with a Green's function approach, we obtain 
\begin{equation}
H_{b 1, l}\left(x_b\right) = \frac{4 \pi}{\left(2 l + 1\right)} \left[\frac{1}{x_b^{l+1}} \int_0^{x_b} dz \, z^{l+2} f_{b 1, l}\left(z\right) + x_b^l \int_{x_b}^{\infty} d z \, z^{-l+1} f_{b 1, l}\left(z\right)\right]
\label{eq:Hb1l}
\end{equation}
and 
\begin{multline}
G_{b 1, l}\left(x_b\right) = - \frac{4 \pi}{\left(4 l^2 - 1\right)} \left[x_b^l \int_{x_b}^{\infty} d z \, z^{-l+3} f_{b 1, l}\left(z\right) - \frac{2l - 1}{2l + 3} x_b^{l + 2} \int_{x_b}^{\infty} dz \, z^{-l+1} f_{b 1, l}\left(z\right) \right. \\ \left.
- \frac{2l - 1}{2l + 3} \frac{1}{x_b^{l + 1}} \int_0^{x_b} dz \, z^{l+4} f_{b 1, l}\left(z\right) + \frac{1}{x_b^{l - 1}} \int_0^{x_b} dz \, z^{l+2} f_{b 1, l}\left(z\right)
\right]
\label{eq:Gb1l}
\end{multline}
as integrals of $f_{b1, l}$, as in
Eqs.~(40) and (45) in Ref.~\cite{RosenbluthPotentials}. 
To find $C_{ab}^{H}$ and $C_{ab}^{G}$ we need $\displaystyle \p H_{b1, l} / \p x_b$ and $\displaystyle \p^2 G_{b1, l} / \p x_b^2$ which are computed by analytically differentiating 
Eqs.~\eqref{eq:Hb1l}-\eqref{eq:Gb1l}.
}
\new{
We evaluate Eqs.~\eqref{eq:Hb1l}-\eqref{eq:Gb1l} and their derivatives replacing $f_{b1,l}$ by each of the polynomials $M_n^k$, using integration endpoints $x_b$ corresponding to each speed grid point for species $a$ normalized to $v_b$. 
For each Legendre mode, the results for $N_x$ polynomials and $N_x$ evaluation points yield a $N_x \times N_x$ matrix which we denote by $R$. 
Thus, the map from distribution for species $b$ (on the speed collocation grid points for species $b$) 
to perturbed Rosenbluth potentials (on the speed collocation grid points for species $a$) is given by the matrix product $R \, Y$. 
The computational expense of these integrations is negligible compared to solving the main linear system of discretized kinetic equations. For most circumstances, the method of evaluating $C_{ab}^{H}$ and $C_{ab}^{G}$ described here gives identical results (to 2 or more decimal places) to the method in \cite{landremanComput}; however we find the method here to yield better convergence at extremely high collisionality.
}

\begin{figure}[!ht]
\begin{center}
\includegraphics[width=1.0\textwidth]{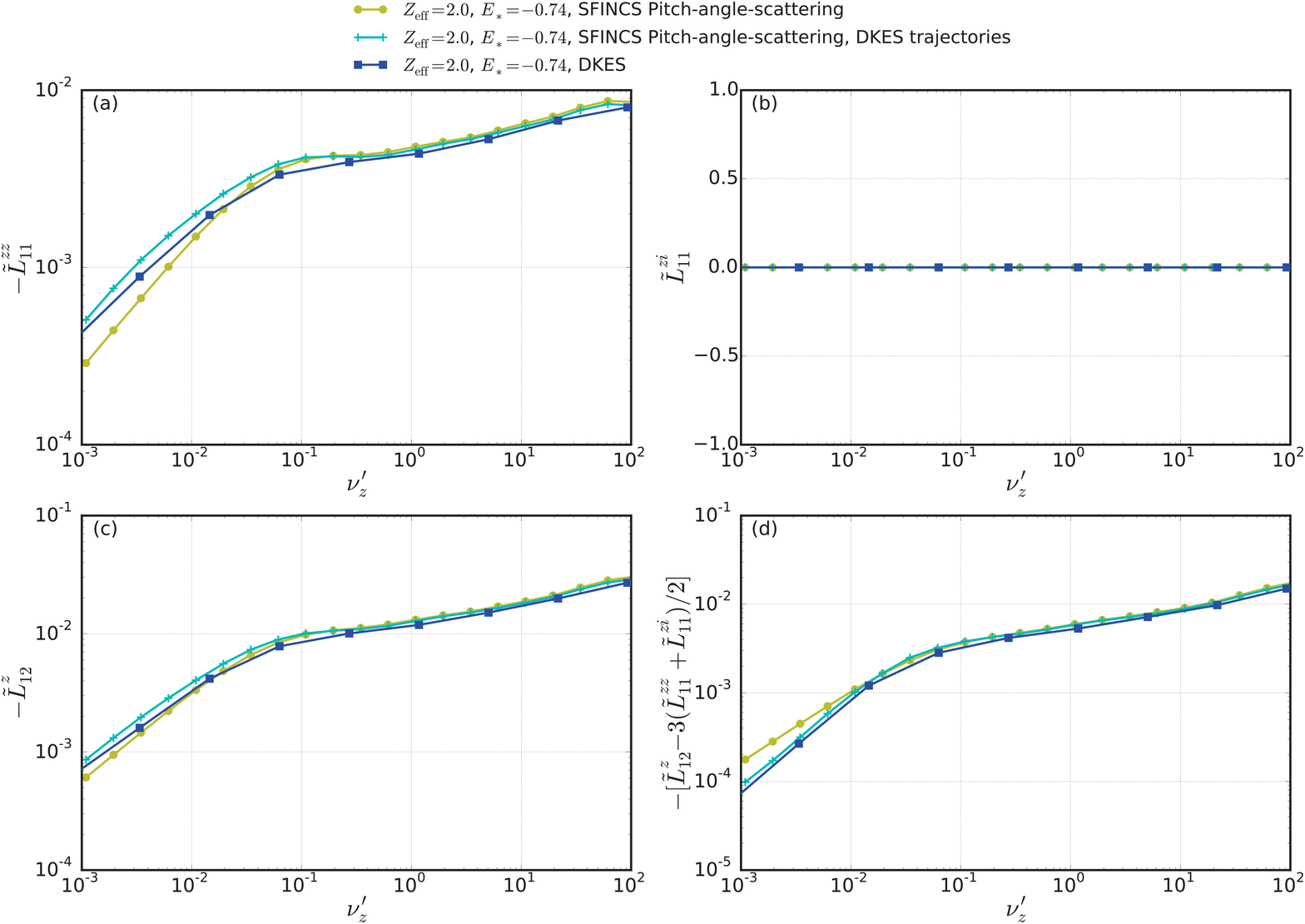}
\end{center}
\caption{
\NewThird{
Carbon ($Z = 6$) transport coefficients $\tilde{L}_{11}^{zz}$ (a), $\tilde{L}_{11}^{zi}$ (b), $\tilde{L}_{12}^{z}$ (c) and the temperature gradient coefficient $\tilde{L}_{12}^{z} - 3 \left(\tilde{L}_{11}^{zz} + \tilde{L}_{11}^{zi}\right)/2$ (d) as functions of normalized collision frequency $\nu_{z}'$ for a W7-X geometry at $E_{\ast} = -0.74$ 
and with $Z_{\mathrm{eff}} = 2.0$. 
\sfincs~computations for pitch-angle scattering with two different models for the particle trajectories (described in Ref.~\cite{landremanSFINCS}) are compared: ``full particle trajectories''    
(\textcolor{ColorZeff1p05Two}{\fullcircleline})
and ``\dkes~particle trajectories''   
(\textcolor{NewCyan}{\plusline}).
Also shown are results from \dkes~(pitch-angle scattering)  
(\textcolor{Blue}{\fullsquareline}).
Note that $\tilde{L}_{11}^{zi}$ in (b) vanish. 
}
}
\label{fig:CarbonTransportCoefficientsW7XZeff2p0withEr20kVPASDKEStraj}
\end{figure}

\NewSecond{
\section{Impurity transport coefficients for pitch-angle scattering models}
\label{sec:appendixPAS}
In Figs.~\ref{fig:CarbonTransportCoefficientsW7XZeff1p05PAS}, \ref{fig:CarbonTransportCoefficientsW7XZeff2p0PAS} and \ref{fig:CarbonTransportCoefficientsW7XZeff2p0withEr20kVPAS} 
we compare the \sfincs~Fokker-Planck-Landau computations of the impurity transport coefficients in Sec.~\ref{sec:TransportCoefficients} to \sfincs~computations with pitch-angle scattering and \dkes~computations (no momentum correction applied afterwards).} 
\NewThird{Moreover, in Fig.~\ref{fig:CarbonTransportCoefficientsW7XZeff2p0withEr20kVPASDKEStraj} we compare \sfincs~pitch-angle scattering computations using ``\dkes~particle trajectories'' to \dkes~computations. It is reassuring to see that when the two different numeric tools use the same collision operator and the same effective particle trajectories, they yield practically the same results. 
We also see that for this particular case the resulting difference from using different models for the particle trajectories is small. 
}

\NewSecond{
Figures~\ref{fig:CarbonTransportCoefficientsW7XZeff1p05PAS}-\new{\ref{fig:CarbonTransportCoefficientsW7XZeff2p0withEr20kVPAS}}~(a) and (c) show that at low collisionality, momentum conservation is unimportant for $\tilde{L}_{11}^{zz}$ and $\tilde{L}_{12}^{z}$. 
This finding is consistent 
with the results of \cite{landremanSFINCS} where a single ion species was analyzed, and is explained as follows. 
In the low-collisionality $1/\nu_{z}'$-regime the radial transport is connected to pitch-angle scattering of helically trapped particles, 
and the dominant physics is captured by the pitch-angle scattering approximation. 
If a radial electric field is present this is also true for the $\sqrt{\nu_{z}'}$-regime. 
The effect of the collisions is mainly to scatter particles across the trapped-passing boundary in velocity space. }

\NewSecond{
In the high-collisionality regime, the difference in $\tilde{L}_{11}^{zz}$ is small between the momentum-conserving linearized Fokker-Planck-Landau operator and the pitch-angle scattering operator at low $Z_{\mathrm{eff}}$, whereas at $Z_{\mathrm{eff}} = 2.0$  and $E_{\ast} = 0$, the pitch-angle-scattering result is a factor of $\sim 5$ larger than the Fokker-Planck-Landau result. 
However, at $Z_{\mathrm{eff}} = 2.0$ and $E_{\ast} = -0.74$ 
the difference is smaller. 
In contrast, for $\tilde{L}_{12}^{z}$ the sign of the coefficient depends crucially on which collision operator is used 
for both values of $Z_{\mathrm{eff}}$. 
This is also verified by the results in Appendix~\ref{sec:appendixB}. 
For both $\tilde{L}_{11}^{zz}$ and $\tilde{L}_{12}^{z}$ the \dkes~curves 
conform reasonably well with the \sfincs~pitch-angle scattering curves. 
}

\NewSecond{
Furthermore, we see that for pitch-angle scattering the ion density gradient coefficient disappears in all collisionality regimes: $\tilde{L}_{11}^{zi} = 0$ (and also $\tilde{L}_{12}^{zi} = 0$), which is 
confirmed by the results 
presented in Appendix~\ref{sec:appendixB} 
and can be understood as follows. 
In the absence of a momentum-conserving term, the impurities only feel collisions with a stationary background. 
In the impurity drift-kinetic equation, there is no information about the density gradient of the main (bulk) ions, 
and consequently the radial impurity flux is independent thereof. 
If momentum is conserved in the collisions, however, the impurities are affected (through the collision operator) by the bulk ion flux along the magnetic field, 
which depends on the ion density gradient. }

\NewSecond{
Finally, Figs.~\ref{fig:CarbonTransportCoefficientsW7XZeff1p05PAS}-\ref{fig:CarbonTransportCoefficientsW7XZeff2p0PAS}~(d) show that at intermediate and high collisionality, 
the absence of momentum correction in the collision operator can lead to transport predictions in the wrong direction for the temperature gradient coefficient. 
In the \sfincs~Fokker-Planck-Landau calculations a temperature screening is typically found (except at very high collisionality for the $Z_{\mathrm{eff}} = 2.0$ case), 
but pitch-angle scattering calculations predict an inward impurity drive. 
However for the calculations at finite $E_{\ast}$ (Fig.~\ref{fig:CarbonTransportCoefficientsW7XZeff2p0withEr20kVPAS}~(d)) both collision models give practically the same result for the temperature gradient coefficient, and a screening is found at all collisionalities. 
}
\NewSecond{
\begin{figure}[!ht]
\begin{center}
\includegraphics[width=1.0\textwidth]{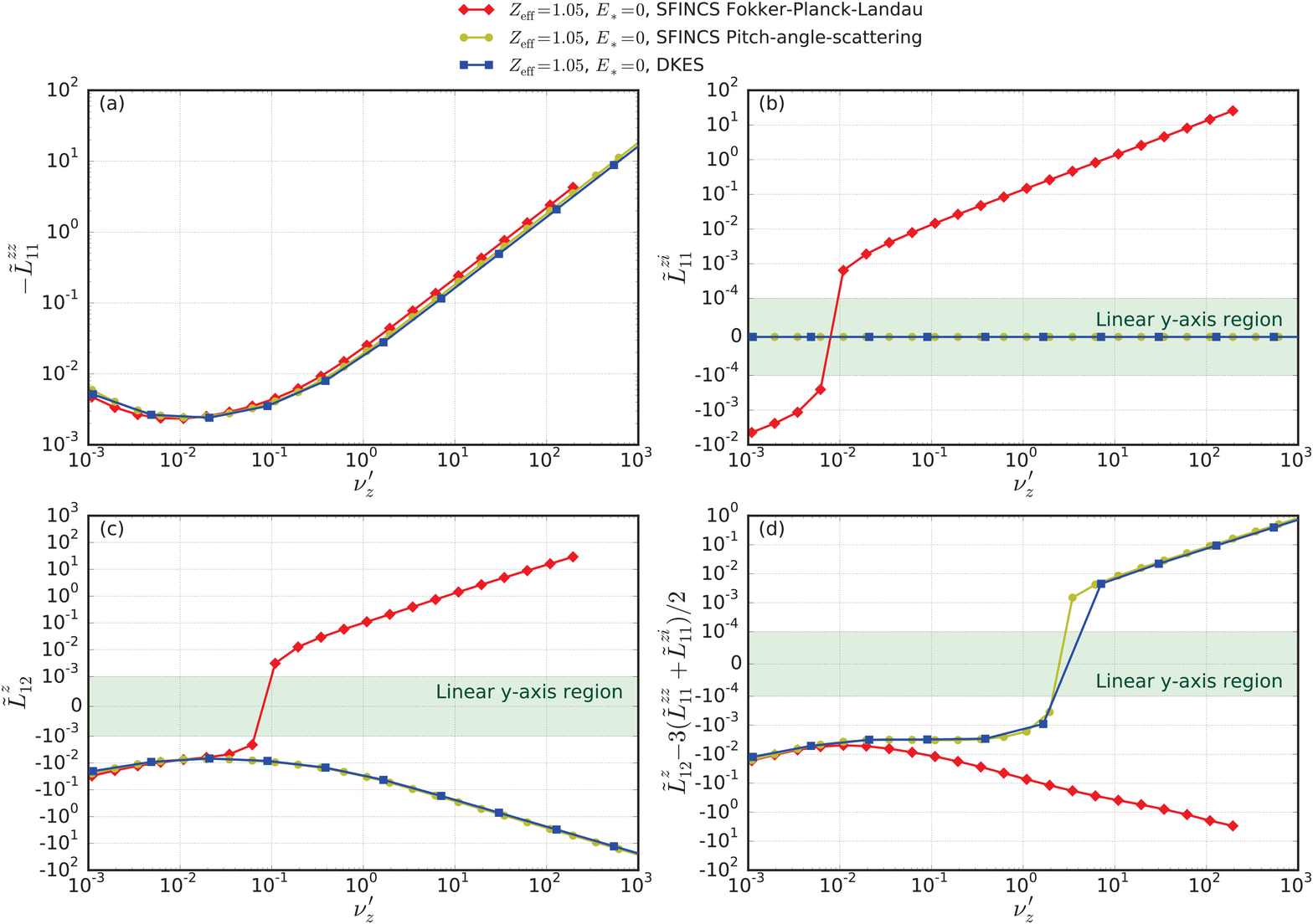}
\end{center}
\caption{
\NewSecond{
Carbon ($Z = 6$) transport coefficients $\tilde{L}_{11}^{zz}$ (a), $\tilde{L}_{11}^{zi}$ (b), $\tilde{L}_{12}^{z}$ (c) and the temperature gradient coefficient $\tilde{L}_{12}^{z} - 3 \left(\tilde{L}_{11}^{zz} + \tilde{L}_{11}^{zi}\right)/2$ (d) as functions of normalized collision frequency $\nu_{z}'$ for a W7-X geometry at $E_{\ast} = 0$ and with $Z_{\mathrm{eff}} = 1.05$. 
\sfincs~computations for two different collision operators are compared: Fokker-Planck-Landau (\textcolor{Red}{\fulldiamondline}) and pitch-angle scattering (\textcolor{ColorZeff1p05Two}{\fullcircleline}). 
Also shown are results from \dkes~(pitch-angle scattering) 
(\textcolor{Blue}{\fullsquareline}). 
Note the double-logarithmic scale in (b)-(d).
} 
}
\label{fig:CarbonTransportCoefficientsW7XZeff1p05PAS}
\end{figure} 
\begin{figure}[!ht]
\begin{center}
\includegraphics[width=1.0\textwidth]{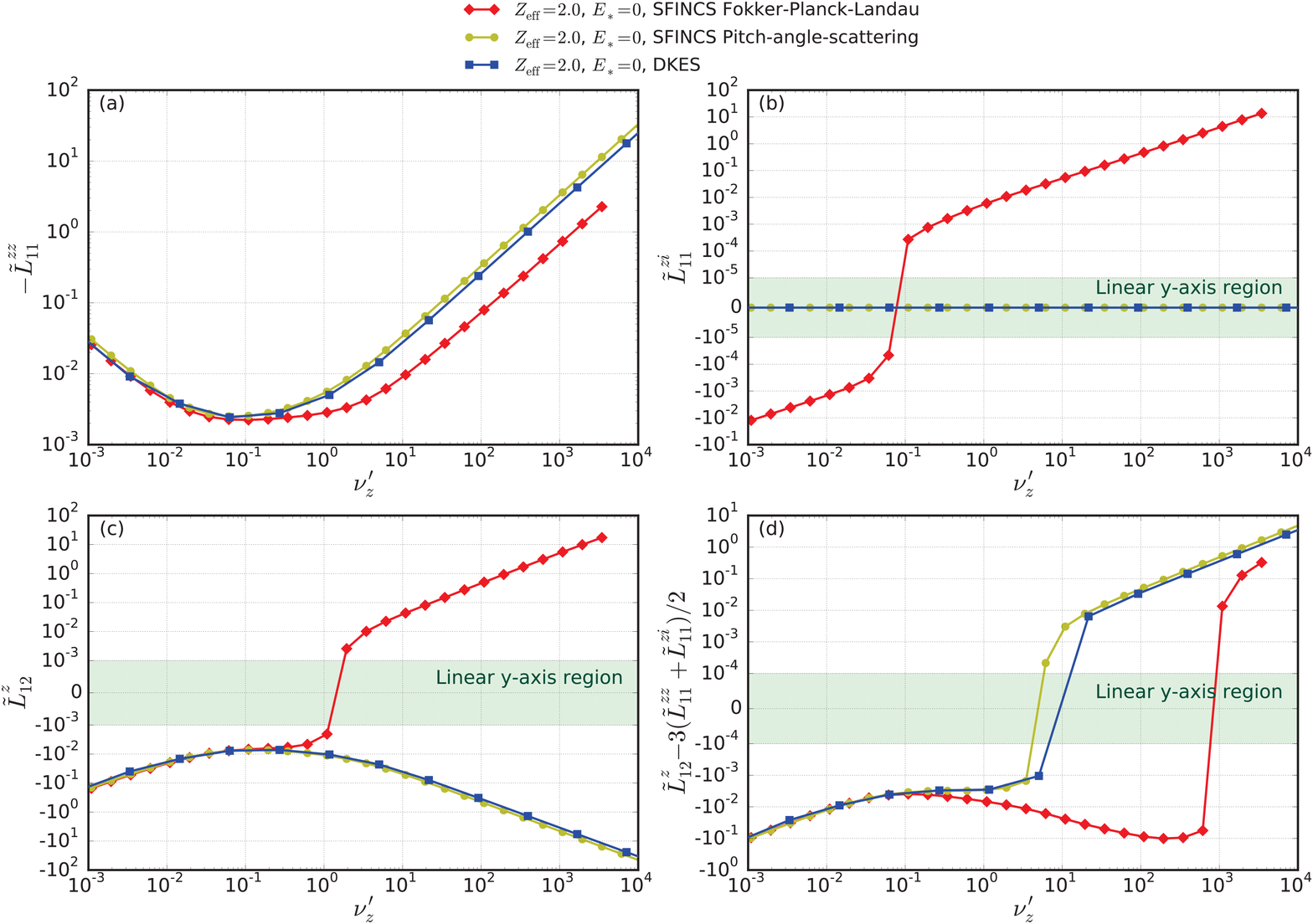}
\end{center}
\caption{
\NewSecond{
Carbon ($Z = 6$) transport coefficients $\tilde{L}_{11}^{zz}$ (a), $\tilde{L}_{11}^{zi}$ (b), $\tilde{L}_{12}^{z}$ (c) and the temperature gradient coefficient $\tilde{L}_{12}^{z} - 3 \left(\tilde{L}_{11}^{zz} + \tilde{L}_{11}^{zi}\right)/2$ (d) as functions of normalized collision frequency $\nu_{z}'$ for a W7-X geometry at $E_{\ast} = 0$ and with $Z_{\mathrm{eff}} = 2.0$. 
\sfincs~computations for two different collision operators are compared: Fokker-Planck-Landau 
(\textcolor{Red}{\fulldiamondline})
and pitch-angle scattering 
(\textcolor{ColorZeff1p05Two}{\fullcircleline}).
Also shown are results from \dkes~(pitch-angle scattering) 
(\textcolor{Blue}{\fullsquareline}).
Note the double-logarithmic scale in (b)-(d). 
}
}
\label{fig:CarbonTransportCoefficientsW7XZeff2p0PAS}
\end{figure}
\begin{figure}[!ht]
\begin{center}
\includegraphics[width=1.0\textwidth]{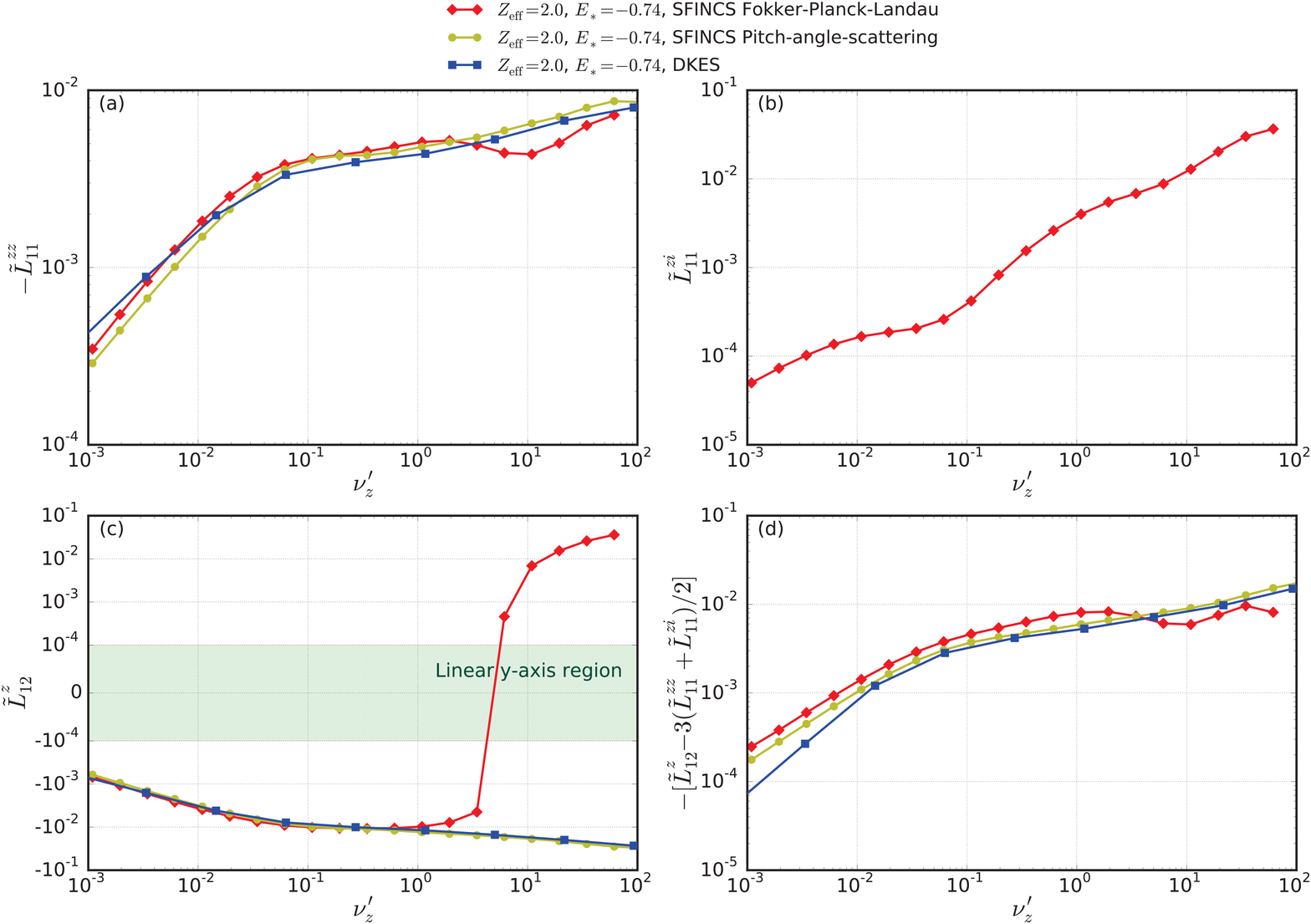}
\end{center}
\caption{
\NewSecond{
Carbon ($Z = 6$) transport coefficients $\tilde{L}_{11}^{zz}$ (a), $\tilde{L}_{11}^{zi}$ (b), $\tilde{L}_{12}^{z}$ (c) and the temperature gradient coefficient $\tilde{L}_{12}^{z} - 3 \left(\tilde{L}_{11}^{zz} + \tilde{L}_{11}^{zi}\right)/2$ (d) as functions of normalized collision frequency $\nu_{z}'$ for a W7-X geometry at $E_{\ast} = -0.74$ 
and with $Z_{\mathrm{eff}} = 2.0$. 
\sfincs~computations for two different collision operators are compared: Fokker-Planck-Landau  
(\textcolor{Red}{\fulldiamondline})
and pitch-angle scattering  
(\textcolor{ColorZeff1p05Two}{\fullcircleline}).
Also shown are results from \dkes~(pitch-angle scattering)  
(\textcolor{Blue}{\fullsquareline}).
Note the double-logarithmic scale in (c) and that the pitch-angle scattering results in (b) vanish.
}
}
\label{fig:CarbonTransportCoefficientsW7XZeff2p0withEr20kVPAS}
\end{figure}
}

\section{Impurity transport coefficients at high collisionality}
\label{sec:appendixB}
In Ref.~\cite{braun} analytic calculations for the impurity transport in the Pfirsch-Schl\"uter regime are presented,  
and from these we can derive expressions for $\new{\tilde{L}_{11}^{zz}}$, $\new{\tilde{L}_{11}^{zi}}$ and $\new{\tilde{L}_{12}^{z}}$. 
\new{The drift-kinetic equation for the first order (in $\rho_\ast$) distribution function $f_{a 1}$ is solved by a subsidiary expansion in the shortness of the
mean free path, $\Delta_i \equiv \lambda_{ii}/L \ll 1$, where $\lambda_{ii} = v_{i}/\nu$ is the ion mean-free-path and $L \sim \nabla^{-1}$ is the plasma dimension. 
The lowest order solution ($f_{a 1}^{(-1)} \sim \Delta_i^{-1}$) is a shifted Maxwellian.} 
The impurity flux is determined from a pressure anisotropy term and an impurity-ion friction term, 
whose relative sizes are $\displaystyle \frac{\mathrm{pressure~anisotropy~term}}{\mathrm{friction~term}} \sim \frac{\lambda_{ii}^2}{L^2 Z^4}$. 
In the short-mean-free-path limit, the pressure anisotropy term can consequently be neglected in transport calculations. 
The friction term is intrinsically ambipolar, and the fluxes are independent of the radial electric field in this limit 
\new{(when the usual drift ordering $v_E \sim \rho_\ast v_i$ is used and $\Ev \times \Bv$-precession is formally excluded)}. 

The coefficients are straightforwardly obtained from Eq.~(2) in Ref.~\cite{braun}, after 
neglecting the pressure anisotropy term.
Note that Ref.~\cite{braun} employs SI-units, and we need to transform the corresponding expressions into Gaussian units to match Eq.~\eqref{eq:ImpFluxSFINCS}. 
The impurity coefficients depend on the geometry-dependent quantity $u$ satisfying
\begin{equation}
\na_{\|} u = \frac{2}{B^2} \left(\mathbf{b} \times \na \psi \right) \cdot \na \ln B,
\label{eq:ParallelCurrentU}
\end{equation}
where $\na_{\|} = \mathbf{b} \cdot \na$ is the gradient along the magnetic field. 
$u$ is proportional to the parallel current divided by $B$. 

Expressions for the impurity transport coefficients in the high collisionality regime, and under the assumption $T_z = T_i = T$, 
are summarized in Eqs.~\eqref{eq:ImpurityTransportCoefficientsPfirschSchluter} and \eqref{eq:Hfunction}.
\begin{equation}
\begin{dcases}
	& \new{\tilde{L}_{11}^{zz}} =  \frac{3 \sqrt{\pi}}{2 Z^{4}} \left( \frac{\beta_{i_1}}{
	\alpha_{i_1} \beta_0 - \alpha_0 \beta_{i_1} 
	}  \right)   \frac{n_i^2}{n_z^2} \frac{m_i^{1/2}}{m_z^{1/2}}  \frac{1 }{G^2} \new{\iota^2} B_0^2   
\, H\left(\psi\right) \,	\nu_{z}', \\ 
	& \new{\tilde{L}_{11}^{zi}} = -Z  \new{\tilde{L}_{11}^{zz}}, \\ 
	& \new{\tilde{L}_{12}^{z}} = - \frac{3 \sqrt{\pi}}{2 Z^{4}}  \left( 
 \frac{
  \frac{5}{2} \left(Z - 1\right) \beta_{i_1} +  \beta_0  
 }{
 \alpha_{i_1} \beta_0 - \alpha_0 \beta_{i_1} 
 }  
 \right)  
 \frac{n_i^2}{n_z^2} \frac{m_i^{1/2}}{m_z^{1/2}} 
 \frac{1}{G^2}  \new{\iota^2} B_0^2 \, H\left(\psi\right) \,  \nu_{z}', 
\label{eq:ImpurityTransportCoefficientsPfirschSchluter} 
\end{dcases}
\end{equation}
\begin{equation}
H\left(\psi\right) = \frac{\left\langle u B^2\right\rangle^2}{\left\langle B^2\right\rangle} - \left\langle u^2 B^2\right\rangle \leq 0.
\label{eq:Hfunction}
\end{equation}
$\alpha_0$, $\alpha_{i_1}$, $\beta_0$ and $\beta_{i_1}$ are coefficients of $f_{i1}^{\left(0\right)}$ expanded in Sonine polynomials, 
with $f_{i1}^{\left(0\right)}$ being the zeroth order term in an expansion in the shortness of the ion mean-free-path 
of the first order (in $\rho_\ast$) 
ion distribution $f_{i1}$. 
The details are given in Ref.~\cite{braun}, and here we have calculated the coefficients for $Z = 6$ and $m_z / m_i = 11.924$. 
For $Z_{\mathrm{eff}} = 1.05$ we obtain  
$\alpha_0 = -24.689$, 
$\alpha_{i_1} = 1.076$, 
$\beta_0 = -2.689$, 
$\beta_{i_1} = 2.857$, 
and for  $Z_{\mathrm{eff}} = 2.0$ we obtain  
$\alpha_0 = -1.741$, 
$\alpha_{i_1} = 0.574$, 
$\beta_0 = -1.435$, 
$\beta_{i_1} = 1.812$. 

\new{From a calculation similar to Ref.~\cite{braun} but only including pitch-angle scattering collisions we obtain the impurity transport coefficients 
}
\begin{equation}
\begin{dcases}
	& \new{\tilde{L}_{11, \mathrm{PAS}}^{zz}} =  4 \mathcal{S}_{Z_{\mathrm{eff}}} \frac{1 }{G^2} \new{\iota^2} B_0^2   
\, H\left(\psi\right) \,	\nu_{z}', \\ 
	& \new{\tilde{L}_{11, \mathrm{PAS}}^{zi}} = \new{\tilde{L}_{12, \mathrm{PAS}}^{zi}} = 0, \\ 
	& \new{\tilde{L}_{12, \mathrm{PAS}}^{z}} =  \left(\mathcal{R}_{Z_{\mathrm{eff}}} + \frac{5}{2}\right) \, \new{\tilde{L}_{11, \mathrm{PAS}}^{zz}}. 
\label{eq:ImpurityTransportCoefficientsPfirschSchluterPAS} 
\end{dcases}
\end{equation}
\new{
Here $\mathcal{R}_{Z_{\mathrm{eff}}}$ and $\mathcal{S}_{Z_{\mathrm{eff}}}$ depend on the impurity content, 
for $Z_{\mathrm{eff}} = 1.05$ they are $\mathcal{R}_{1.05} = - 1.04549$, $\mathcal{S}_{1.05} = 1.82082$ 
and for $Z_{\mathrm{eff}} = 2.0$ they are $\mathcal{R}_{2.0} = - 1.14447$, $\mathcal{S}_{2.0} = 0.329224$.
}

\NewSecond{
The high-collisionality asymptotes for the W7-X case in Sec.~\ref{sec:TransportCoefficients} are plotted in Figs.~\ref{fig:CarbonTransportCoefficientsW7XZeff1p05PfirschSchluter} and \ref{fig:CarbonTransportCoefficientsW7XZeff2p0PfirschSchluter}, 
and compared to \sfincs~calculations at $E_{\ast} = 0$ 
with both full linearized Fokker-Planck-Landau collisions and pitch-angle scattering 
(note that since \sfincs~includes $\Ev \times \Bv$-precession whereas the analytic high-collisionality calculations do not, a comparison at finite $E_{\ast}$ is not meaningful). 
We find that the \sfincs~results conform well with the analytic predictions. 
}
\NewSecond{
\begin{figure}[!ht]
\begin{center}
\includegraphics[width=1.0\textwidth]{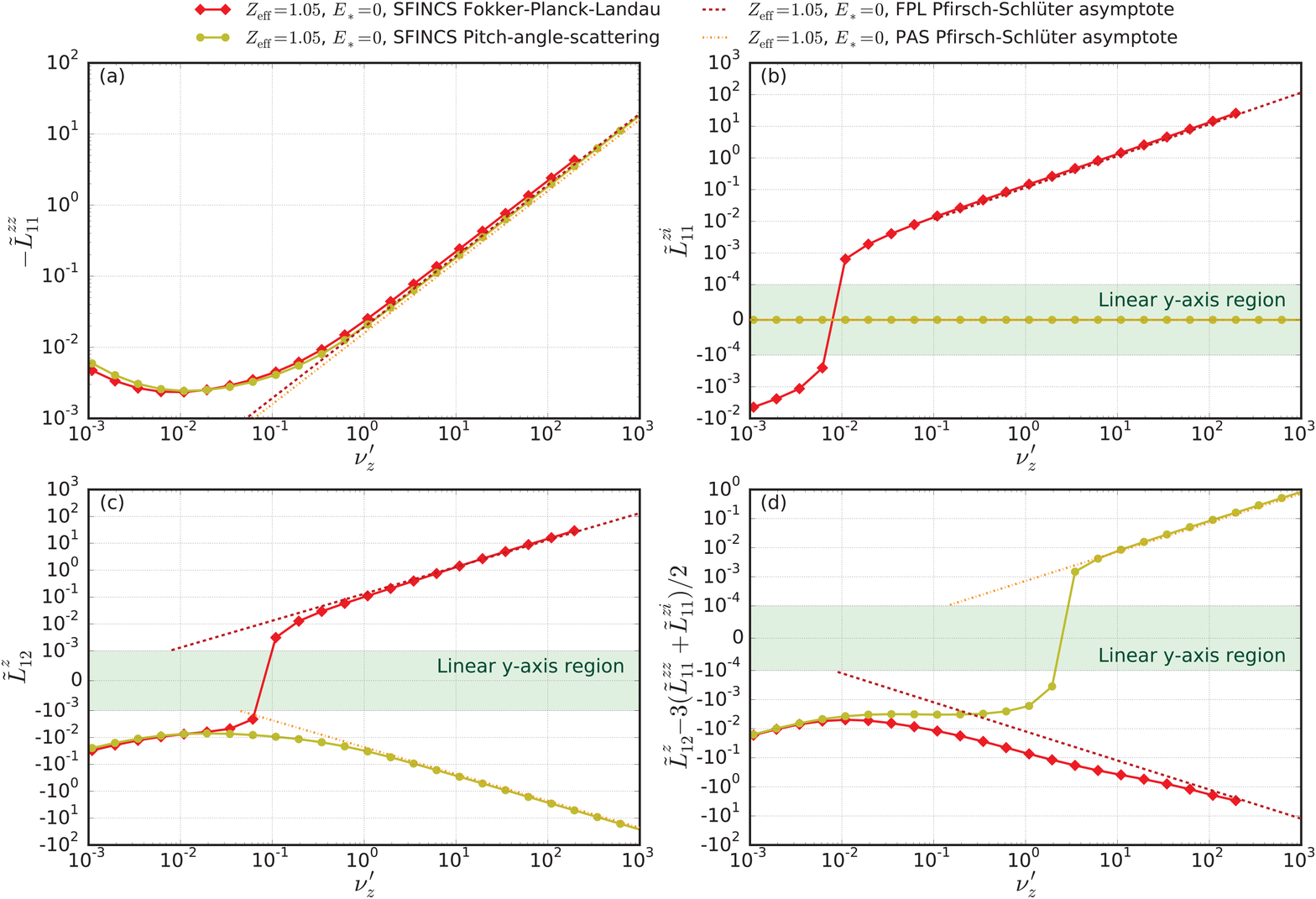}
\end{center}
\caption{
\NewSecond{
Carbon ($Z = 6$) transport coefficients $\tilde{L}_{11}^{zz}$ (a), $\tilde{L}_{11}^{zi}$ (b), $\tilde{L}_{12}^{z}$ (c) and the temperature gradient coefficient $\tilde{L}_{12}^{z} - 3 \left(\tilde{L}_{11}^{zz} + \tilde{L}_{11}^{zi}\right)/2$ (d) as functions of normalized collision frequency $\nu_{z}'$ for a W7-X geometry at $E_{\ast} = 0$ and with $Z_{\mathrm{eff}} = 1.05$. 
\sfincs~computations for Fokker-Planck-Landau collisions (\textcolor{Red}{\fulldiamondline}) and pitch-angle scattering (\textcolor{ColorZeff1p05Two}{\fullcircleline}) 
are compared to the analytic high-collisionality limits for Fokker-Planck-Landau collisions (\textcolor{ColorZeff1p05Five}{\dashed}) and pitch-angle scattering  (\textcolor{ColorZeff1p05Six}{\chain}).
Note the double-logarithmic scale in (b)-(d). 
}
}
\label{fig:CarbonTransportCoefficientsW7XZeff1p05PfirschSchluter}
\end{figure} 
\begin{figure}[!ht]
\begin{center}
\includegraphics[width=1.0\textwidth]{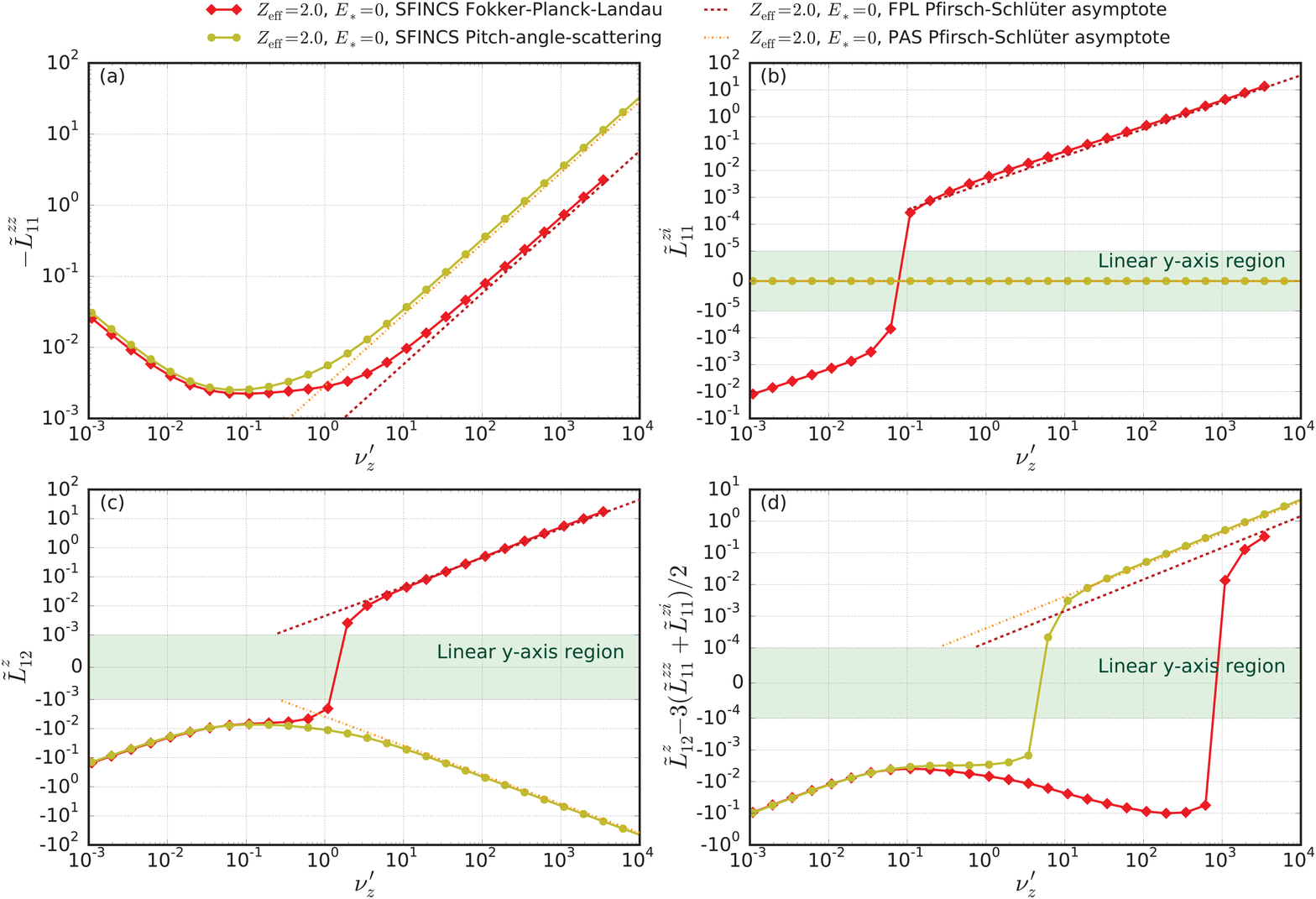}
\end{center}
\caption{
\NewSecond{
Carbon ($Z = 6$) transport coefficients $\tilde{L}_{11}^{zz}$ (a), $\tilde{L}_{11}^{zi}$ (b), $\tilde{L}_{12}^{z}$ (c) and the temperature gradient coefficient $\tilde{L}_{12}^{z} - 3 \left(\tilde{L}_{11}^{zz} + \tilde{L}_{11}^{zi}\right)/2$ (d) as functions of normalized collision frequency $\nu_{z}'$ for a W7-X geometry at $E_{\ast} = 0$ and with $Z_{\mathrm{eff}} = 2.0$. 
\sfincs~computations for Fokker-Planck-Landau collisions (\textcolor{Red}{\fulldiamondline}) and pitch-angle scattering (\textcolor{ColorZeff1p05Two}{\fullcircleline}) 
are compared to the analytic high-collisionality limits for Fokker-Planck-Landau collisions (\textcolor{ColorZeff1p05Five}{\dashed}) and pitch-angle scattering  (\textcolor{ColorZeff1p05Six}{\chain}).
Note the double-logarithmic scale in (b)-(d). 
}
}
\label{fig:CarbonTransportCoefficientsW7XZeff2p0PfirschSchluter}
\end{figure}
}

\bibliographystyle{unsrt}

\end{document}